\documentclass[%
10pt,
twocolumn,
superscriptaddress,
floatfix,
preprintnumbers,
nofootinbib,
amsmath,amssymb,
aps, prd,
]{revtex4-2}

\pdfoutput=1

\usepackage{lmodern}
\usepackage{graphicx}
\usepackage{siunitx}
\usepackage{physics}
\usepackage{dcolumn}
\usepackage{bm}
\usepackage{color}
\usepackage[dvipsnames,table]{xcolor}
\usepackage[colorlinks=true, pdfstartview=FitV,  breaklinks=true]{hyperref}
\hypersetup{urlcolor=BlueViolet, citecolor=Plum, linkcolor=PineGreen}
\usepackage{booktabs}

\usepackage{lettrine}
\usepackage{Zallman}

\begin{document}

\preprint{ \includegraphics[width=0.4\textwidth]{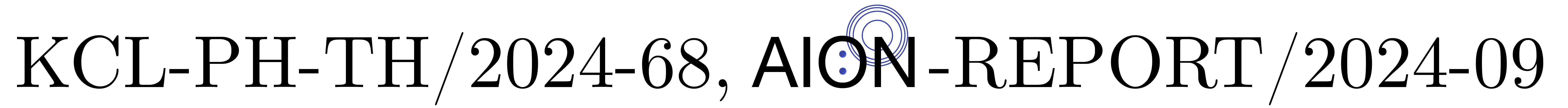}  }

\title{Clear skies ahead: characterizing atmospheric \\ gravity gradient noise for vertical atom interferometers}

\date{\today}

\author{John~Carlton}
\email{john.carlton@kcl.ac.uk}
\thanks{co-first authors.}
\affiliation{Physics Department, King’s College London, Strand, London, WC2R 2LS, UK}

\author{Valerie~Gibson}
\affiliation{Cavendish Laboratory, University of Cambridge, J.J.~Thomson Avenue, Cambridge CB3 0HE, UK}

\author{Tim~Kovachy}
\affiliation{Department of Physics and Astronomy and Center for Fundamental Physics,\\ Northwestern University,
Evanston, IL, USA}

\author{Christopher~McCabe}
\affiliation{Physics Department, King’s College London, Strand, London, WC2R 2LS, UK}

\author{Jeremiah~Mitchell}
\email{jm2427@cam.ac.uk}
\thanks{co-first authors.}
\affiliation{Cavendish Laboratory, University of Cambridge, J.J.~Thomson Avenue, Cambridge CB3 0HE, UK}

\begin{abstract}
Terrestrial long-baseline atom interferometer experiments are emerging as powerful tools for probing new fundamental physics, including searches for dark matter and gravitational waves. In the frequency range relevant to these signals, gravity gradient noise (GGN) poses a significant challenge. While previous studies for vertical instruments have focused on GGN induced by seismic waves, atmospheric fluctuations in pressure and temperature also lead to variations in local gravity. In this work, we advance the understanding of atmospheric GGN in vertical atom interferometers, formulating a robust characterization of its impact. We evaluate the effectiveness of underground placement of atom interferometers as a passive noise mitigation strategy. Additionally, we empirically derive global high- and low-noise models for atmospheric pressure GGN and estimate an analogous range for atmospheric temperature GGN. To highlight the variability of temperature-induced noise, we compare data from three prospective experimental sites. Our findings establish atmospheric GGN as comparable to seismic noise in its impact and underscore the importance of including these effects in site selection and active noise monitoring for future experiments.
\end{abstract}

\maketitle
\flushbottom









\section{Introduction}
\label{sec:introduction}

\lettrine{T}{he} advent of large-scale quantum sensors is ushering in a new era of scientific precision. 
Among these, atom interferometers (AIs) are a promising class that have demonstrated exquisite precision in tests of fundamental physics.
AIs at the scale of \SIrange{1}{10}{m} have already been used to measure fundamental constants~\cite{Rosi:2014,Morel:2020dww, Parker_2018} and test foundational aspects of quantum mechanics~\cite{Arndt_2014, Manning:2015cta, Bassi:2012bg, Vowe:2022mzm} and gravity~\cite{Xu:2019vlt,Asenbaum_2020,Overstreet:2022} including  proposed tests of general relativity~\cite{Roura:2018cfg, Zych:2011hu}. They have also constrained models of dark energy~\cite{Hamilton:2015zga, Elder:2016yxm, Burrage:2014oza, Sabulsky:2018jma} and `fifth' forces~\cite{Biedermann_2015,Panda:2024}. 
Furthermore, they have demonstrated practical applications in inertial sensing, navigation, and absolute measurements of gravity~\cite{Dubetsky:2006,Dickerson:2013,Tino:2021,Stray:2022}. 

Several groups are planning or constructing very long-baseline detectors in both vertical (V) and horizontal (H) orientations.
This includes AION (V)~\cite{Badurina:2019hst}, ELGAR (H)~\cite{Canuel:2020elgar}, MAGIS-100 (V)~\cite{MAGIS-100:2021etm}, MIGA (H)~\cite{Canuel:2017rrp}, and ZAIGA (V\&H)~\cite{Zaiga}. 
There are also plans for space-based AIs, such as AEDGE~\cite{AEDGE:2019nxb} and STE-QUEST~\cite{Aguilera:2013uua}. 
Horizontal configurations feature two baselines in a geometry similar to LIGO~\cite{LIGOScientific:2014pky}, while vertical detectors have a single baseline aligned with the Earth's gravitational field.

These longer-baseline AIs have been proposed as sensors to search for ultralight dark matter (ULDM)~\cite{Graham:2015ifn, Geraci:2016fva, Arvanitaki:2016fyj, Badurina:2021lwr, Badurina:2021rgt, DiPumpo:2022muv, Badurina:2023wpk} and gravitational waves (GWs)~\cite{Dimopoulos:2007cj, Dimopoulos:2008sv, Yu:2010ss, Chaibi:2016dze, Graham:2017pmn, Loriani:2018qej, Schubert:2019ycf, Badurina:2024rpp}. 
Their peak sensitivity occurs in the decihertz frequency band, a previously unexplored region corresponding to ULDM masses of \SIrange{1e-16}{1e-13}{\eV}, and a `mid-band' range (\SIrange{0.1}{3}{Hz}) complementary to LIGO-Virgo-KAGRA~\cite{KAGRA:2013rdx} and LISA~\cite{LISA:2017pwj}.

As terrestrial detectors increase in size and sensitivity, they become increasingly limited by gravity gradient noise (GGN), also known as Newtonian noise~\cite{Saulson:1984, Hughes:1998}.
GGN arises from fluctuations in the gravitational potential, which can result from local density changes caused by seismic waves, atmospheric disturbances, or anthropogenic activity. 
This noise is a fundamental limiting factor in the low-frequency sensitivity of {\it all} high precision metrology devices below approximately \SI{10}{\hertz}. 

For that reason, the laser interferometric gravitational wave community has made significant efforts to understand the couplings between natural sources of gravitational fluctuations.
For laser interferometers, a precise discrimination between an incoming gravitational wave that alters the light travel time along the baseline and density fluctuations that accelerate the interferometer mirrors is required.
These mirrors serve as the test masses in the gravitational field.
The gravitational couplings between the mirrors and Earth are expected to become the limiting terrestrial factor as local vibration noise suppression advances.
Exploring these noise sources is an active field, and understanding and mitigating GGN from all sources will be a requirement for low-frequency sensitivity to gravitational waves for the next generation of observatories, such as Cosmic Explorer~\cite{CosmicExplorer:2019} and the Einstein Telescope~\cite{EinsteinTelescope:2010zz}.

GGN arising from {\it seismic activity} has been investigated for both horizontal and vertical AIs. 
Although the GGN source models between laser and atomic interferometers are the same, the responses of the detectors differ significantly~\cite{Baker:2012ck, Vetrano:2013qqa, Harms:2013raa}, and require new calculations. 
Building on the extensive work done by the laser interferometry community, a preliminary characterization of seismic GGN was carried out by the MIGA experiment~\cite{Canuel:2016}.
The proposed configuration of the MIGA detector is analogous to a laser interferometer, i.e., two horizontal baselines measuring a differential phase shift in both baselines. 
Recent work by MIGA has also modeled seismic GGN from oceanic swell closest to the detector's location~\cite{Bertoldi:2024}. 
Studies have also explored seismic GGN in vertical configurations, such as MAGIS-100~\cite{Mitchell:2022}, and the use of multigradiometry to suppress GGN in AI dark matter searches~\cite{Badurina:2023}. 
Explorations of the unseen impact of humans and other transient animal movements have also been carried out to understand these gravitational noise sources~\cite{Carlton:2023ffl, Badurina:2024rpp}.

Developments in {\it atmospheric sourced} GGN have been advanced by Creighton~\cite{Creighton:2008} and Harms~\cite{Fiorucci:2018had, Harms:2019} in the context of the laser interferometers, LIGO and Virgo, in addition to recent work for the upcoming Einstein Telescope experiment~\cite{Brundu:2022mac}.
Interest in these low-frequency effects has again been sparked by the next generation of GW instruments aiming to push sensitivity below \SI{10}{Hz}. 
In atom interferometry, atmospheric-sourced GGN has received less attention compared to seismic noise induced GGN.
MIGA was the first to translate pressure wave GGN as an anomalous strain for horizontal AIs in GW searches~\cite{MIGAconsortium:2019efk}. 
However, a similar characterization for vertical AIs remains absent. Horizontal interferometers, such as MIGA, would typically expect atmospheric noise to be uncorrelated between test masses. Underground vertical setups would instead feel correlated noise along the baseline, fundamentally altering how these backgrounds are modeled.

This article presents the first robust characterization of atmospheric GGN for vertical baseline AIs. 
We focus on two key sources: pressure fluctuations from infrasound waves (below \SI{20}{Hz}) and temperature fluctuations modeled as effective density variations.
Our analysis focuses on the mid-band frequency range, specifically \SIrange{0.01}{1}{Hz}, where AIs are most sensitive to GWs and ULDM, and explores mitigation strategies such as underground detector placement.
 
We address two key challenges in characterizing thermal turbulence for AIs: the absence of dedicated AI modeling and the highly model-dependent nature of low-frequency thermal turbulence. 
While turbulence models typically converge to the same power law at higher frequencies, this convergence breaks down below a few Hz. 
We therefore introduce a model-agnostic method for analyzing the impact of turbulence noise on vertical baseline AIs that accommodates different turbulence models and site-specific empirical data.

Utilizing global limits of infrasound noise from atmospheric science and temperature fluctuation bounds from international monitoring data, we establish upper and lower limits for expected AI phase noise in terrestrial detectors. 
These bounds serve multiple purposes: guiding site selection for next-generation interferometers, similar to Peterson's noise models for seismic reference~\cite{Peterson:1993};
 identifying frequency ranges where atmospheric noise dominates over seismic GGN; and informing potential mitigation strategies.

The rest of the paper is structured as follows. 
In Sec.~\ref{sec:atmo-ggn}, we introduce the fundamentals of atmospheric GGN for atom interferometers, including the basic operating principles and distinguishing different noise sources. 
We present detailed models of both pressure fluctuations from infrasound waves and temperature turbulence.
In Sec.~\ref{sec:atmoggn-results}, we derive the response of vertical AIs to these noise sources
we present our results on the impact of atmospheric GGN, including depth dependence of the noise, establishment of global noise limits from monitoring data, analysis of site-specific variations through a detailed case study, and comparison with seismic GGN.
In Sec.~\ref{sec:discussion} we discuss the implications of our work, exploring the challenges faced to modeling atmospheric GGN, potential mitigation strategies, and the dependence of our work on the AI sequence used.
Sec.~\ref{sec:conclusion} presents a summary of our findings and outlook for future work.

\section{Atmospheric gravity gradient noise}
\label{sec:atmo-ggn}

Before discussing atmospheric pressure and temperature noise, we briefly review the most pertinent
operational details of AI experiments, and provide a high-level discussion of how noise couples through gravitational potential perturbations. For a more detailed review, see, for example~\cite{Canuel:2006zz, Hu:2019, Buchmueller:2023nll}.

\subsection{Review of atom interferometry}

\begin{figure}[!t]
    \centering
    \includegraphics[width=\columnwidth]{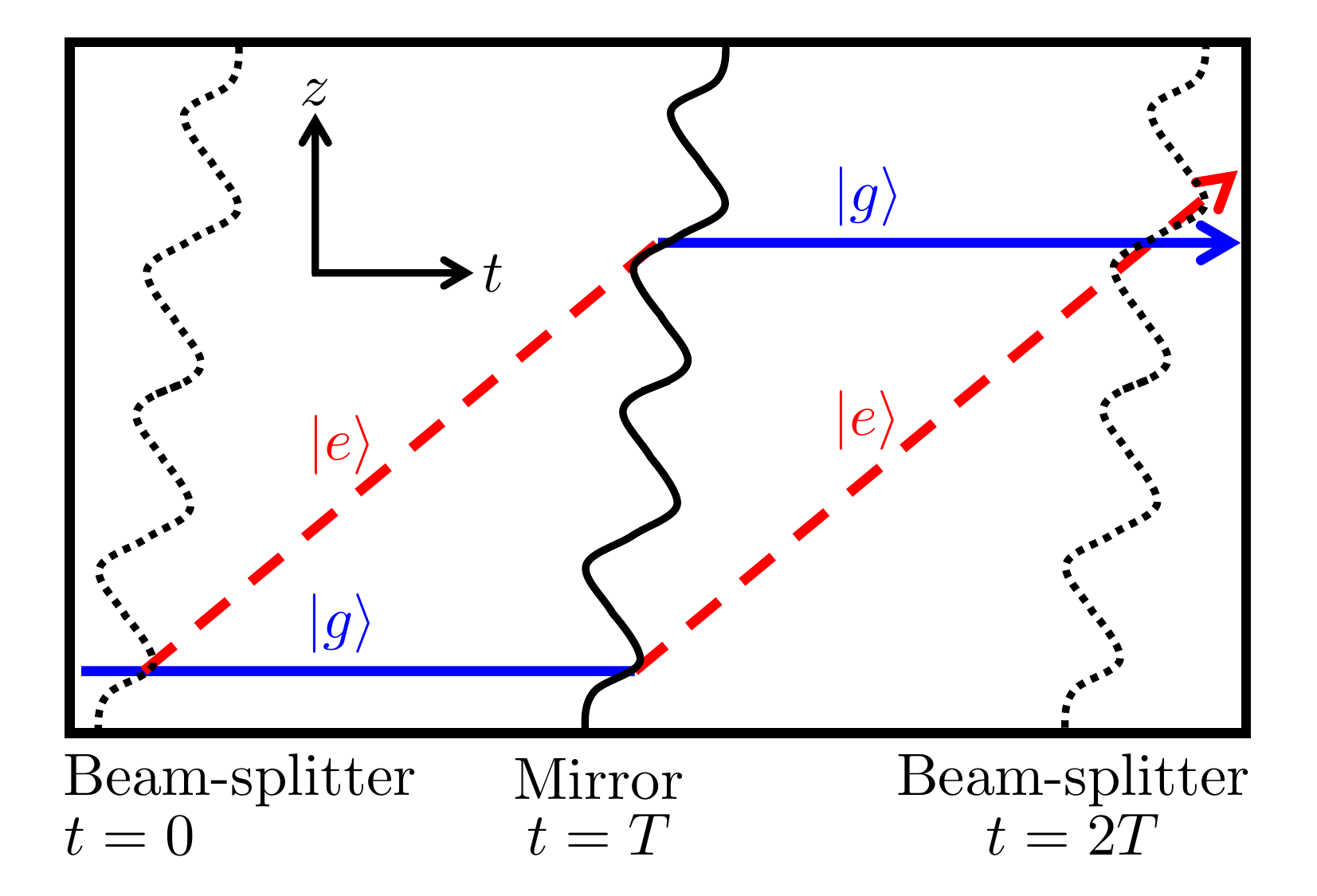}
    \caption{Schematic of a vertical atom interferometer (AI) in free space. The atoms and laser pulses travel in the $z$-direction, aligned with the Earth's gravitational field. A population of atoms prepared in the ground state $|g\rangle$, split into a superposition with an excited state $|e\rangle$ by a beam-splitter ($\pi/2$) pulse. The atoms are redirected by a central mirror ($\pi$) pulse a time $T$ later before interfering with a final beam-splitter ($\pi/2$) pulse at a time $2T$.}
    \label{fig:AI_schema}
\end{figure}

Light pulse AI’s operate using the quantum mechanical principles of superposition and interference. Fig.~\ref{fig:AI_schema} shows a schematic of the interferometer sequence in a vertical AI. Quantum states following different spacetime trajectories accumulate phase offsets if passing through distinct backgrounds e.g., fluctuating gravitational fields. Starting with a single quantum state and delocalizing the state in spacetime, then recombining at a later time can lead to interference from the accumulation of phase between the superposition of states. This AI operation can be realized with a physical system of ultracold neutral atoms. Each atom in an ensemble will undergo the interferometry sequence and acquire phase dependent on the trajectory of the superposed energy or momentum states of the atoms. The splitting and recombination of the atomic wavefunction is achieved through precise atom-light interactions and the atoms follow a free-fall trajectory between light pulses acquiring phase offsets from the trajectory and interactions with the light. The final interference and phase shift of the ensemble of neutral atoms is read out through the difference of populations in the quantum states.

In this paper, we consider a vertical baseline AI in a gradiometer configuration: a configuration of two or more spatially-separated AIs that are referenced by common laser sources. There are many AI experiments that fit this design, from small field-deployable and table-top instruments, to long-baseline detectors of \SI{10}{m}--\SI{15}{m}; such as, VLBAI~\cite{Schlippert:2024}, rubidium equivalence principle tower and planned Strontium tower at Stanford~\cite{Kovachy:2015}, and the Wuhan \SI{10}{m} tower~\cite{Zhou:2011}. This is also the design for the \SI{100}{m} to kilometer scale detectors of MAGIS-100, AION, and ZAIGA experiments. While detectors at the \SI{10}{m} scale have been built in university laboratories in basements and at the surface level, a vertical baseline at advanced scales of \SI{100}{m} or more necessitates underground access or mine shafts. 

\begin{figure}[!t]
    \centering
    \includegraphics[width=\columnwidth]{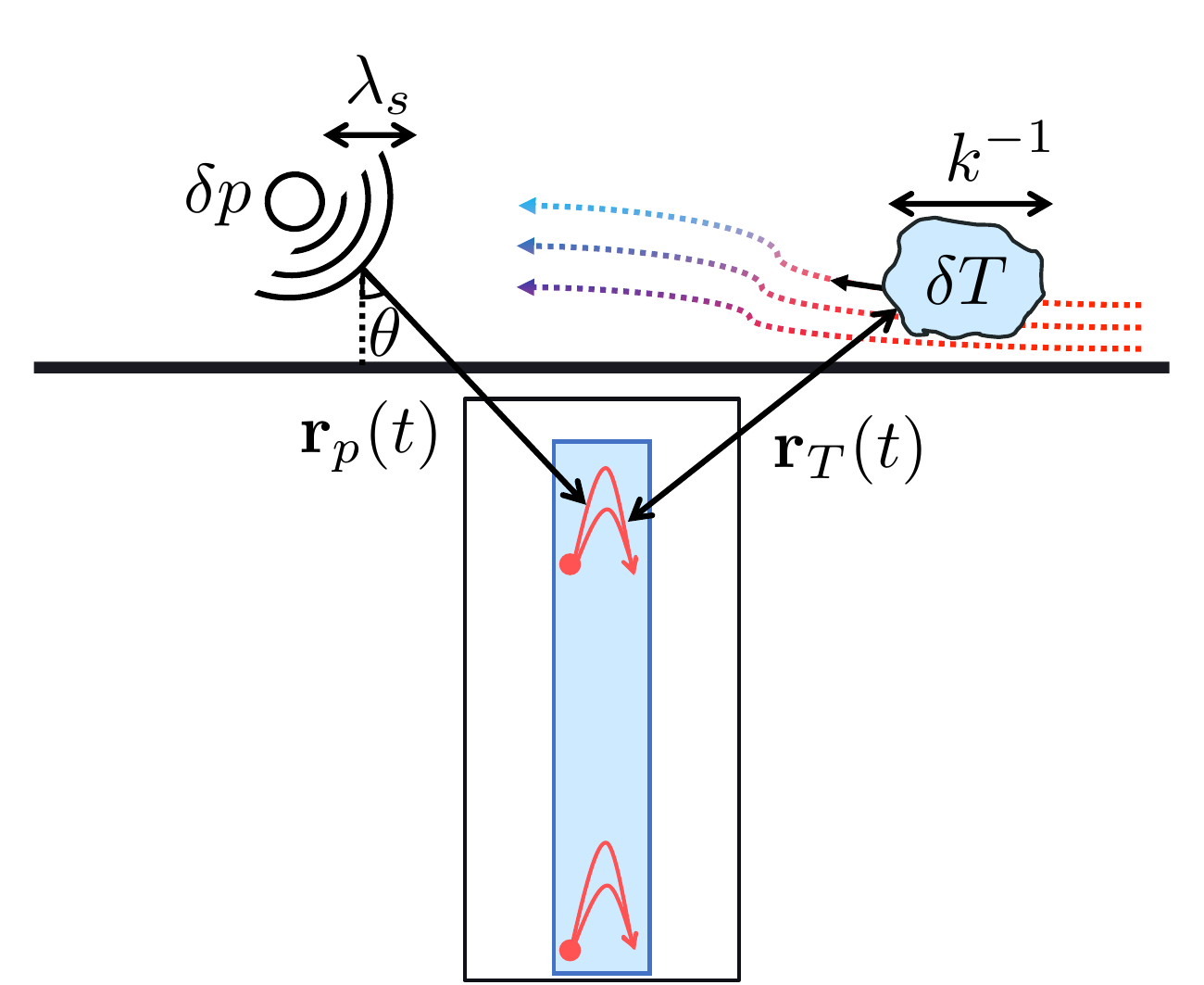}
    \caption{Schematic of atmospheric noise sources affecting an underground AI experiment. Above the surface, pressure ($\delta p$) and temperature ($\delta T$) variations generate density fluctuations that create perturbations in the gravitational potential. Each noise source has a characteristic length scale, which corresponds to the frequency of the noise: pressure waves by their wavelength~$\lambda_s$ and temperature perturbations by their inverse wavenumber~$k$. Pressure waves propagate at an incident angle~$\theta$ relative to the surface normal. The AI operates in a gradiometer configuration with two sources (red circles) launching atoms in vertical trajectories (red parabolas) at different depths, with different proximity to the noise sources represented by arrows.
   }
    \label{fig:noises}
\end{figure}

In a gradiometer configuration, the differential measurement compares an upper AI near Earth's surface with a lower AI at depth.
This means the uppermost AI in the gradiometer configuration is nearer the Earth's surface and more susceptible to atmospheric effects compared with the deeper AI. We have depicted this configuration in Fig.~\ref{fig:noises}, where the red circles and parabolic trajectories represent the upper and lower AI as parts of the gradiometer configuration. The two sources of GGN that we consider are pressure variations from infrasound waves and density fluctuations from 
temperature fluctuations.
Infrasound is depicted as a traveling pressure wave characterized by an incident angle of the wave with respect to the surface normal and the wavelength of sound. Temperature fluctuations source thermal eddies, which are characterized by the inverse wavenumber, corresponding to the size of the eddy.

The effect of GGN on an AI is long range and enters through the propagation of the atoms through the fluctuating gravitational potential. Density fluctuations lead to perturbations of the gravitational potential. The quantum states of the atoms (i.e., the test masses) in our interferometer moving through this perturbed potential acquire a phase shift caused by the different accelerations from the gravity potential gradients. The number of sources generating such density perturbations is potentially vast. The gravitational potential perturbations take the general form
\begin{equation}
    \label{eq:2}
    \delta\phi_g(\vb{r},t) = -G_N\int_{\mathcal{H}}\dd V' \frac{\delta\rho(\vb{r}',t)}{\abs{\vb{r}-\vb{r}'}},
\end{equation}
where $G_N$ is Newton's gravitation constant, $\delta\rho$ is the density fluctuation, and the integral is taken over some domain set by the effect in which we are interested. The change to local gravitational acceleration experienced by an atom is then found by taking the gradient of the gravitational potential. For a vertical AI in an underground shaft, we model the region as an infinite half-space $(\mathcal{H})$ with isotropic and homogeneous layers of fluid above the surface and solid Earth below the surface.  The half-space surface is defined at $z=0$.
More complex domains would include a stratified multi-layer Earth model and vertically varying fluid layers. 

Local vibrations can enter the noise spectrum of an AI through various routes during atom preparation, launch, and interferometry. However, these forms of local technical noise have been studied in detail previously~\cite{MAGIS-100:2021etm} and can largely be suppressed through engineering solutions and gradiometry. In contrast, GGN enters through the propagation of the atoms through the fluctuating gravitational potential and requires different mitigation strategies.

\subsection{Atmospheric pressure fluctuations as noise}
\label{subsec:pressure-fluctuations}

The gravitational response of laser interferometers to infrasound was first investigated as a strain noise source by Saulson~\cite{Saulson:1984} and Creighton~\cite{Creighton:2008}, where they found that it had a limited impact for frequencies above a few Hz. This work has also been summarized and expanded more recently by Harms~\cite{Harms:2019}. 
Infrasound covers a large frequency range from sub-\si{mHz} to tens of \si{Hz} and can be a source of GGN through the density fluctuations of the air creating a local gravitational gradient. In this range we are most concerned with ambient infrasound fluctuations as these would act like a constant noise background.\footnote{Transient infrasound can be sourced by earthquakes, volcanoes, thunderstorms, meteors, and explosions~\cite{Averbuch:2020,Evers:2008,Edwards:2007,Lin:2007}. These sources can propagate as infrasound waves for many hundreds to thousands of \si{km} along the Earth/atmosphere surface and higher up in the atmospheric layers~\cite{Hedlin:2012,Drob:2003} and also contribute to the overall GGN through seismic density fluctuations.} 

The ambient field in the surface boundary layer, the part of the atmosphere that feels direct effects from the Earth's surface and which ranges from a few meters to kilometers above the surface depending on local meteorology, contains many infrasound waves and may not be fully known at a given moment. Therefore, we restrict our investigation to plane infrasound waves incident with the Earth's surface. Pressure waves are similar to seismic P-waves, or compressional longitudinal waves, by creating density fluctuations throughout the transport medium. 

In the frequency band we consider in this work, \SIrange{0.01}{1}{Hz}, infrasound propagation is an adiabatic process for sufficiently small deviations from the mean ambient pressure~\cite{Pierce:1989}.\footnote{This decouples the effects of pressure and temperature perturbations in the atmosphere for our frequency range. In general there may be higher order effects that we neglect in this work.} The adiabatic index for air, $\gamma$, allows us to relate these pressure fluctuations to density fluctuations of the local atmosphere as 
\begin{equation}
    \label{eq:3}
    \frac{\delta\rho}{\rho_{0}}=\frac{\delta p}{\gamma p_{0}},
\end{equation}
where $p_{0}$ and $\rho_{0}$ are the mean pressure and density, respectively. 
We assume the values $\rho_0 = \SI{1.3}{\kilo\gram\per\metre\cubed}$, $p_{0} = \SI{1013}{mbar}$ and $\gamma = 1.4$.

We follow the approach of Creighton and Harms and model the infrasound as a plane wave propagating in a half-space with the wave reflecting from the Earth's surface.
While this model has some limitations, which we will discuss further below, it
allows us to investigate vertically and horizontally incident waves, which through Eq.~\eqref{eq:3}, transport density fluctuations that modify the
gravitational potential experienced by the atom clouds. The plane wave model is supported by the theory of infrasound propagation in the atmosphere leading to multiple reflections between the upper-atmosphere layers and the Earth's surface~\cite{Hedlin:2012}. The waveguide nature of the atmospheric layers allows for long range propagation~\cite{Drob:2003}. The actual ambient field can be constructed from a Fourier synthesis of a broadband infrasound background. The surface boundary layer of the atmosphere is also chaotic and turbulent as will be seen in Sec.~\ref{subsec:temp-fluctuations}. We will see that for low frequencies, this atmospheric sourced gravitational background can be significant for AIs.


\begin{figure*}[!t]
    \begin{minipage}{0.49\textwidth}
    \centering
    \includegraphics[width=\textwidth]{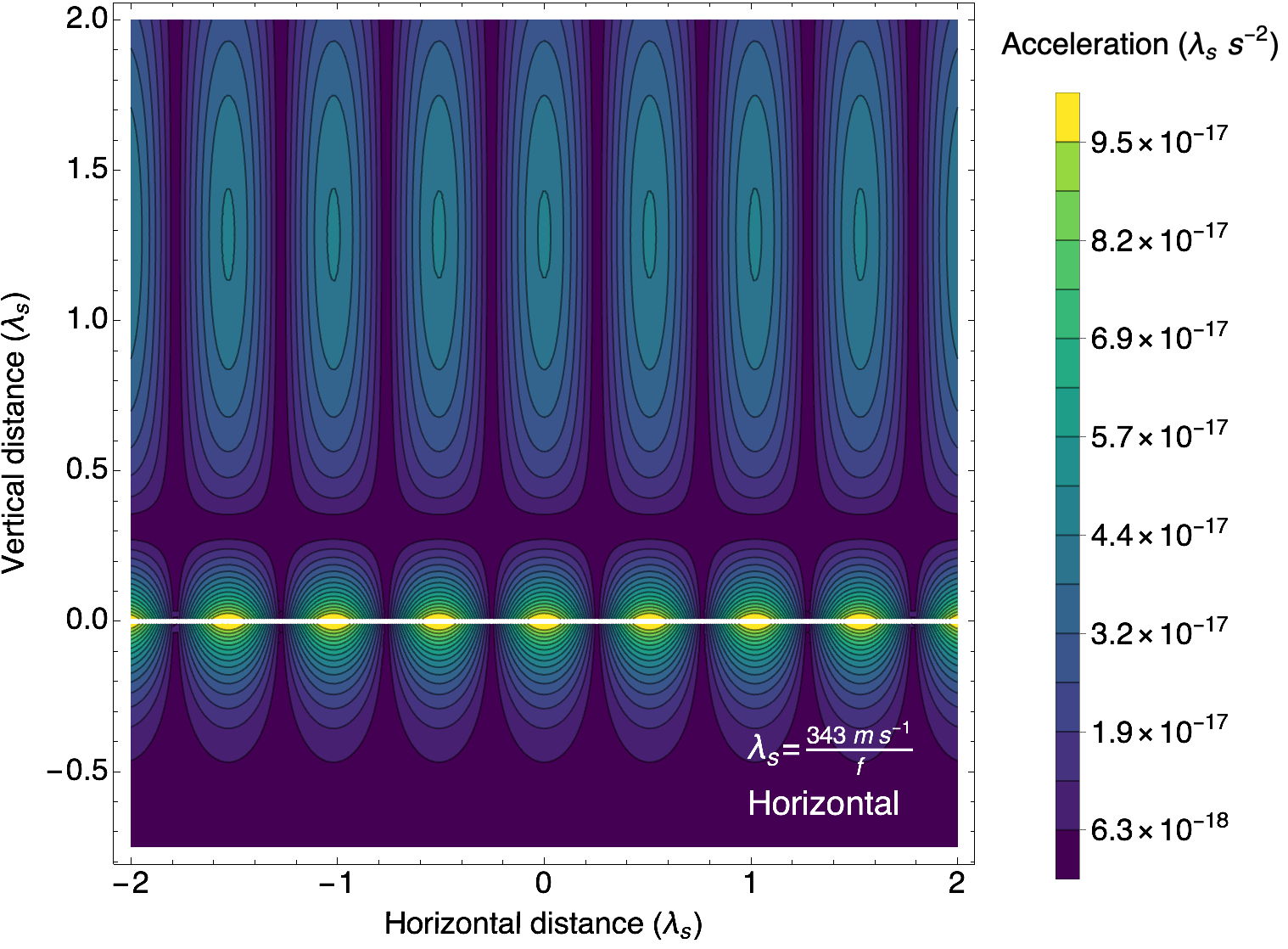}
    \end{minipage}%
    \hfill
    \begin{minipage}{0.49\textwidth}
    \centering
    \includegraphics[width=\linewidth]{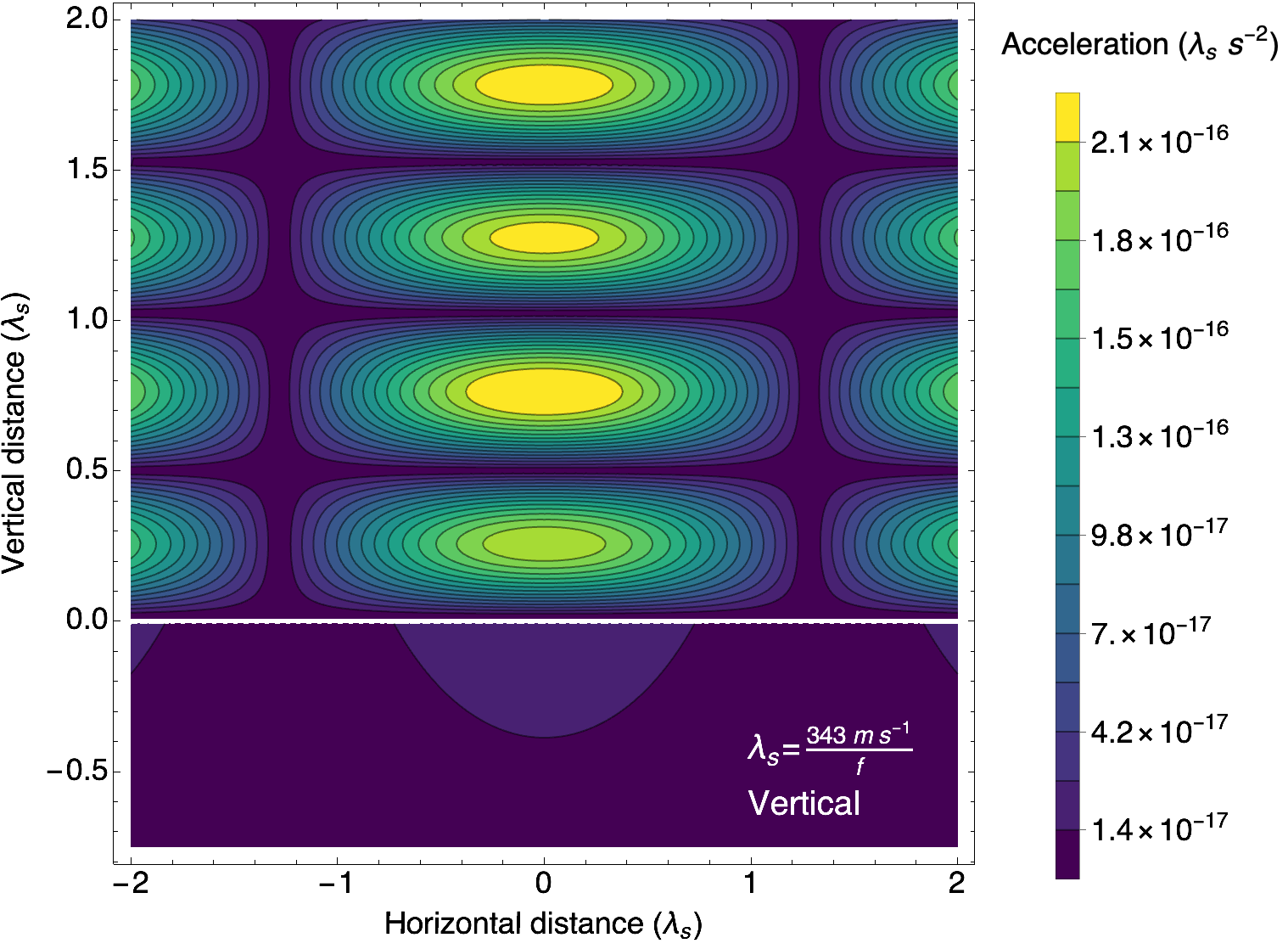}
    \end{minipage}
    \caption{Contour plots of plane wave pressure fluctuation generated vertical acceleration over the halfspace with the Earth's surface at $z=0$ and distances and amplitude in units of the wavelength of the infrasound wave. Left: A nearly horizontally incident plane wave, $\theta = 7\pi/16$. Right: A nearly vertically incident plane wave, $\theta = \pi/16$. Assumed background mean pressure $p_{0} = \SI{1013}{mbar}$ and a fluctuation amplitude $\delta p(\omega) = \SI{e-3}{mbar}$~\cite{Bowman:2005}.}
    \label{fig:pressure-accel}
\end{figure*}

Mathematically, the pressure wave is modeled as
\begin{equation}
\begin{aligned}
    \label{eq:4}
    \delta p(\vb{r}, t) = \int & \frac{{\rm d}\omega}{2\pi}~\delta p(\omega) \exp{i(\vb*{k_{\varrho}} \vdot \vb*{\varrho} - \omega t)} \\
    &\quad  \times \qty(\exp{i k_z z} + \exp{-i k_z z}),
\end{aligned}
\end{equation}
where $\delta p(\omega)$ is the frequency dependent amplitude, working in cylindrical coordinates: $\vb*{\varrho} = (\varrho \cos(\phi),\varrho\sin(\phi),0)$, $\vb*{k_{\varrho}} = (k_\varrho  \cos(\phi'), k_\varrho \sin(\phi'), 0)$, $k_\varrho = k\sin(\theta)$ and $k_z = k\cos(\theta)$ are the horizontal and vertical wavevectors, with~$\theta$ the incident angle with respect to the surface normal ($0<\theta<\pi/2$). Combining Eqs.~\eqref{eq:3} and~\eqref{eq:4} we compute the integral in Eq.~\eqref{eq:2} resulting in a gravitational potential perturbation. We define a coupling constant~$\alpha$, 
\begin{equation}
    \label{eq:5}
    \alpha = G_{N}\frac{\rho_0}{\gamma p_0},
\end{equation}
with the gravitational potential
\begin{equation}
    \begin{aligned}
  \label{eq:6}
  \delta\phi_g(\vb{r}_{0},t) &= -\alpha\int\frac{{\rm d}\omega}{2\pi} \exp{i(\vb*{k_{\varrho}}\vdot \vb*{\varrho}_{0} - \omega t)}\delta p(\omega)\\
  &\times \int_{\mathcal{H}}\dd V \frac{\exp{i k_{\varrho} \varrho \cos(\phi-\phi')}\qty(e^{i k_{z} z}+e^{-i k_{z} z})}{\sqrt{\varrho^{2} + (z - z_{0})^{2}}},\\
  &= \frac{4\pi \alpha}{k^{2}}\int\frac{{\rm d}\omega}{2\pi}~\exp{i(\vb*{k}_{\varrho}\vdot\vb*{\varrho}_{0} - \omega t)}\delta p(\omega)\\
  &\times \bigg(e^{-k_{\varrho}\abs{z_{0}}}(2\Theta(z_{0}) - 1)-2\cos(k_{z} z_{0})\Theta(z_{0})\bigg),
  \end{aligned}
\end{equation}
where $(x_0,z_0)$ are the horizontal and vertical position of the AI, $\Theta(z_0)$ is the Heaviside theta function for the vertical position of the AI with respect to the surface ($\Theta(0) \equiv 0$), $k$ is the pressure wave wavevector magnitude such that $k^2 = k_\varrho^2 + k_z^2$, and $\omega$ is the frequency of the incident plane wave. 

The corresponding vertical acceleration for a single Fourier mode is
\begin{equation}
\begin{aligned}
\label{eq:7}
    g_z ={}& -\partial_z (\delta\phi_g), \\ 
    ={}& \frac{4\pi \alpha }{k^2 z_0} \;e^{- k_\varrho \abs{z_0}}\cos(\omega t - k_\varrho x_0)\delta p(\omega)\\
    &\times\bigg[k_\varrho \abs{z_0}(2\Theta(z_0)-1) \\
    & \qquad - 2 k_z z_0 \;e^{k_\varrho \abs{z_0}} \sin(k_z z_0) \Theta(z_0)\bigg].
\end{aligned}
\end{equation}
Illustrative acceleration contour plots in units of the wavelength of the infrasound wave are shown in Fig.~\ref{fig:pressure-accel} for a nearly horizontal ($\theta = 7\pi/16$) and nearly vertical ($\theta = \pi/16$) incident wave.\footnote{Purely horizontal and vertical infrasound waves, corresponding to $\theta = \pi/2$ and $\theta = 0$, respectively, would amount to an infrasound wave perfectly confined to the surface in the horizontal case, and in the vertical case a plane wave with infinite transverse extent, and there would be no effects below the surface.}
To illustrate an average level of noise, we choose $\delta p(\omega) =2\pi\delta(\omega-\omega_0) \times \SI{e-3}{mbar} $~\cite{Bowman:2005}, where the delta function picks out a single Fourier mode. For reference, infrasound at a frequency of \SI{1}{Hz} corresponds to a wavelength of \SI{343}{m}.

These contour plots show the vertical acceleration component experienced by a test mass (an atom), in contrast to the horizontal acceleration component, which is of primary interest to laser interferometers measuring strain along a horizontal baseline. However, the AI baseline is aligned with the Earth's gravitational field and is sensitive to vertical acceleration perturbations.\footnote{The phase shift is also sensitive to transverse accelerations from the Coriolis effect, at the level of \SIrange{0.4}{5}{rad}~\cite{Dickerson:2013}. These fluctuations can also affect the contrast of the interferometry measurement by coupling to the initial kinematics of the atom cloud as another source of noise limiting overall sensitivity. Mitigation strategies for this are well known for smaller AI~\cite{Dickerson:2013}, with plans for longer baseline AIs under development~\cite{Glick:2024}.}  Above the surface, test masses map out these gradients and, depending on the overall scale of the trajectory, can see a significant influence, whereas underground for both the horizontal and vertical pressure waves there is a much simpler structure to account for. 

In Fig.~\ref{fig:pressure-accel}, we depict only a single Fourier frequency.
A more complex stochastic model could be constructed by assuming a random admixture of wave modes and frequencies and random or correlated phases if interested in the properties of incoherent and coherent ambient noise. 
This may lead to an unpredictable sea of accelerations acting on the AI. For higher frequencies, above a few Hz, the horizontal components of the pressure wave can lead to a spatially varying gravitational potential impacting the transverse motion of the atoms in the AI complicating the background noise further.

Our infinite half-space model assumes isotropic and homogeneous layers of fluid above the surface.
For atmospheric infrasound modeling, assuming isotropy is not sufficient as it could lead to un-physical averaging down of the ambient noise through destructive interference. Modeling and experiment of long-range infrasound propagation show clear anisotropy~\cite{Drob:2003,Evers:2008} and for different locations on Earth the wind is narrowly distributed with a preferential angular direction, influenced by global and regional wind circulation~\cite{Larsen:2022}. This plays a role in guiding infrasound propagation and in advecting temperature fluctuations, i.e.,\ thermal eddies. For predictive modeling and data interpretation the frequency distribution of the wind is critical although it will not account for directionality of transient events.

We have also assumed complete reflection of the infrasound wave such that density fluctuations do not propagate through the surface into the Earth. However, there will still be a fluctuation in the gravitational potential below the surface with an exponentially decaying amplitude. This is seen by setting $\Theta(z_0)$ to zero in Eq.~\eqref{eq:6}. This assumption does not hold completely for sufficiently energetic sources and can lead to air-coupled Rayleigh waves~\cite{Novoselov:2020,Lognonne:2016,Edwards:2007,Haskell:1951}. The energy coupling admittance of infrasound absorption by the Earth has been measured up to 2\% in experimental studies using co-located infrasound and seismic sensor arrays~\cite{Novoselov:2020,Edwards:2007}. This coupling depends strongly on the surface properties and the velocity profile of the shallow subsurface, specifically if the velocity of shear waves below the surface is of the same order as the speed of sound in air~\cite{Langston:2004}. Air-guided waves can propagate along the surface leading to seismically induced GGN~\cite{Satari:2022wna}. Seismic GGN has been previously explored in the context of atom interferometry~\cite{Canuel:2017rrp,Mitchell:2022,Badurina:2023}. 

Surface experiments have shown a percentage of energy can transfer into seismo-acoustic surface waves and transform into seismic P-waves~\cite{Novoselov:2020,Langston:2004}. This effect is dependent on the composition of the local subsurface. To fully remove the assumption of perfect infrasound reflection, we would need to model the coupling between the solid Earth and fluid atmosphere including the horizontal stratification as done in reference~\cite{Averbuch:2020}. This would provide a more complete ambient field with which to model the gravity gradient noise in general and the response of the atom interferometer.

In this work, we will not investigate these surface effects, or secondary air-coupled Rayleigh waves, as the simplified model of plane waves propagating in a half-space is sufficient to capture the effects of horizontally propagating infrasound. This allows us to set expected noise level limits when we characterize the impact on AIs in Sec.~\ref{sec:atmoggn-results}.

\subsection{Atmospheric temperature fluctuations as noise}
\label{subsec:temp-fluctuations}

We now change focus to discuss noise induced by temperature perturbations in the atmosphere, the biggest contribution coming from eddies in the surface boundary layer. These structures are produced through turbulent mixing and can be thought of as pockets of warm or cool air relative to the ambient temperature. On sufficiently brief timescales, these pockets of air can be thought of as `frozen', since they do not evolve in time and thus act as masses when advected (transported by wind) past a measuring station. The modeling of atmospheric temperature fluctuations is complex but has been historically simplified by working in the frequency domain. We follow this approach in this section to derive the temperature noise spectrum. We will then apply the results to a vertical AI experiment in Sec.~\ref{sec:atmoggn-results}.

As with infrasound, atmospheric temperature perturbations are proportional to density fluctuations~\cite{Creighton:2008, Harms:2019}
\begin{equation}
    \frac{\delta\rho}{\rho_0} =\frac{\delta T}{T_0} .
\end{equation}
From this and Eq.~\eqref{eq:2}, we define a coupling~$\beta$
\begin{equation}
    \beta = G_N\frac{\rho_0}{T_0}, \label{eq:betadef}
\end{equation}
where $T_0$ is the mean air temperature surrounding the detector. We assume the value $T_0 = \SI{300}{\kelvin}$.

\subsubsection{Power spectral density}

For a vertical atom interferometer located underground, the primary noise source will be the turbulent behavior of the atmospheric boundary layer. While the atmosphere is split into distinct layers with different properties, the main source of turbulent noise will come from the surface boundary layer. We characterize this layer as extending from the Earth's surface to a height~$L_O$, known as the Obukhov length. It is calculated according to~\cite{Brundu:2022mac}
\begin{equation}
    L_O = \frac{T_0 u_*^2}{\kappa_\mathrm{VK} g T_*},
    \label{eq:obukhov}
\end{equation}
where $u_*$ is the friction velocity, $\kappa_\mathrm{VK}$ is the Von Kármán constant, $g$ is the acceleration due to gravity, and $T_*$ is the friction temperature. Both the friction velocity and temperature act as a scaling for the vertical logarithmic wind and temperature profiles in the surface layer. The dominant method of turbulent production in the surface layer comes from wind shear, while above a height $L_O$, buoyancy effects become dominant and the wind and temperature profiles tend to a constant. Typically the Obukhov length is on the order of a few hundred meters. However, it is highly variable in both time and space. Characterizing the atmospheric noise in terms of these parameters thus allows us to minimize the location dependence of our analysis.

\begin{figure}[!t]
    \centering
    \includegraphics[width=\columnwidth]{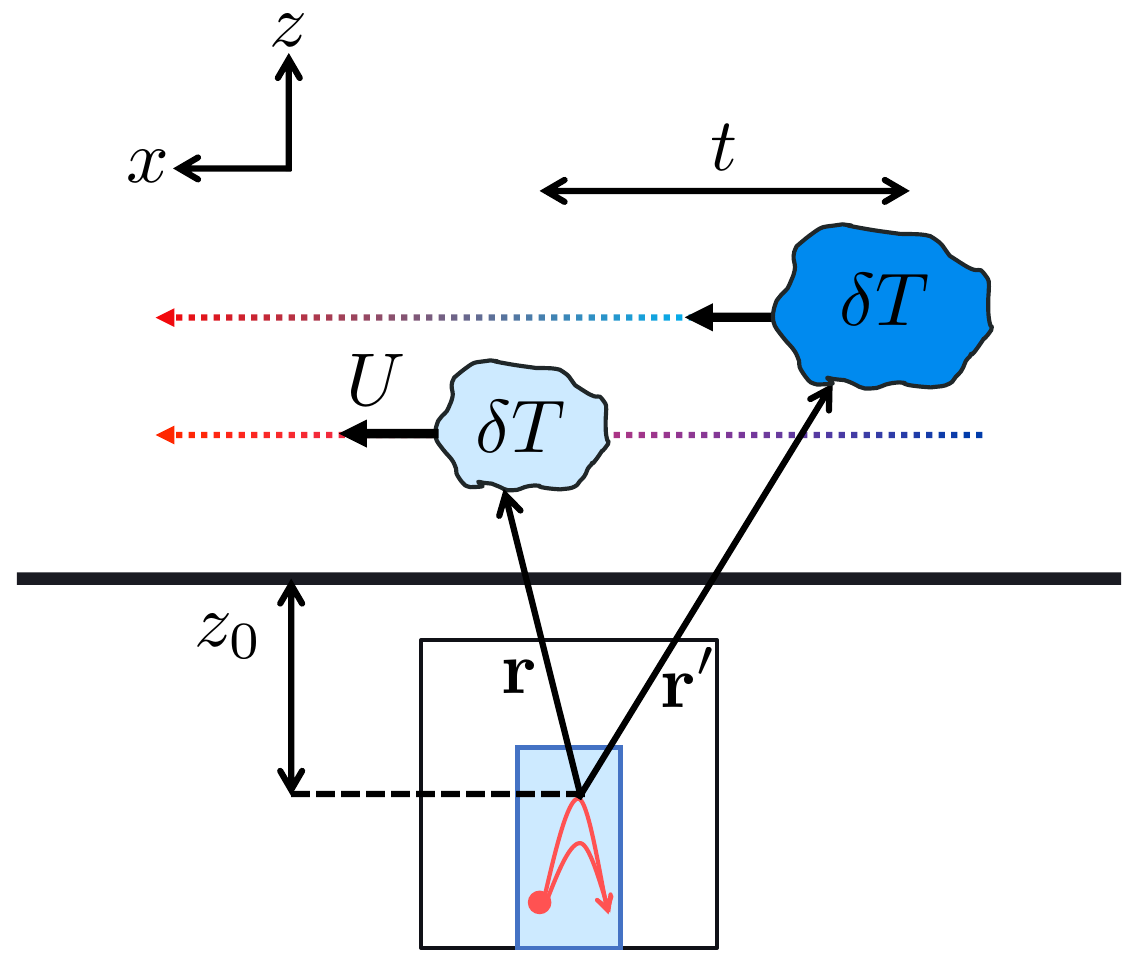}
    \caption{Schematic of frozen temperature perturbations advected past an atom interferometer. The eddies are advected by wind at mean speed $U$ along two horizontal streamlines at distances $\bf{r}$ and $\bf{r}^\prime$ from the atoms. The closest approach is $z_0$. The perturbations are separated in time by $t$. Integrating over the spatial and temporal distribution of the frozen eddies gives the power spectrum of temperature GGN.}
    \label{fig:temp_correlation}
\end{figure}

Describing the turbulent temperature field of the surface layer in the time domain is an intractable problem. Fig.~\ref{fig:temp_correlation} is a schematic of how we model the temperature field coupling to the atoms in an interferometer. Along parallel streamlines, temperature perturbations are advected past the detector at different times and positions. We evaluate the temperature spectrum as a function of frequency, defined as the Fourier transform of the temperature field autocorrelation at two different points in space and time~\cite{Creighton:2008}
\begin{equation}
    S_T(f; \mathbf{r}, \mathbf{r}^\prime) = \int \mathrm{d}t \; \langle \delta T(t, \mathbf{r})\delta T(0, \mathbf{r}^\prime) \rangle e^{2\pi i f t}.
    \label{eq:tempspec}
\end{equation}
This provides a function that describes the spectrum of temperature perturbations for distances $\mathbf{r},\mathbf{r}^\prime$. With the coupling factor $\beta$ defined in Eq.~\eqref{eq:betadef}, we can map this to the spectrum of gravity perturbations. What remains is to integrate the field over all space, using a Green's function $G(\mathbf{r})$ to describe the spatial separation between the perturbation and atoms. Overall this gives the power spectral density of the temperature GGN~\cite{Creighton:2008,Brundu:2022mac}
\begin{equation}
    S_g(f) = \beta^2\int \mathrm{d}t\mathrm{d}^3r \mathrm{d}^3r^\prime  G(\mathbf{r}) G(\mathbf{r}^\prime)   \langle \delta T(t, \mathbf{r})\delta T(0, \mathbf{r}^\prime) \rangle e^{2\pi i f t}.
    \label{eq:sg1}
\end{equation}

Characterizing the temperature spectrum is essential for understanding these fluctuations as a source of GGN and depends on the model of turbulence we use. Assuming the turbulence is homogeneous and isotropic, the spatial dependence of $S_T(f;\mathbf{r},\mathbf{r}^\prime)$ entirely relies on $\mathbf{r}-\mathbf{r}^\prime$. With this assumption, we can convert the spectrum to a wavevector spectrum at a single point in time
\begin{equation}
    \Phi(k) = \int \mathrm{d}^3 r \langle \delta T(0, \mathbf{r})\delta T(0, 0) \rangle e^{-i\mathbf{k} \cdot  \mathbf{r}}.
    \label{eq:wavevecspec}
\end{equation}

To evaluate the time dependence, we employ Taylor's frozen turbulence hypothesis~\cite{Taylor_1938}. We assume eddies are advected by the wind without evolving in time as they are observed by a measuring station. This approximation is useful for simplifying models of atmospheric noise as the frozen eddies can effectively be modeled as test masses in the vicinity of a detector. Each moves along straight streamlines in the $x$~direction, moving with the same average wind speed $U$. Thus the time dependence comes from shifting to the frame of the wind such that $x\rightarrow x+U t$. Combining this with the wavevector spectrum in Eq.~\eqref{eq:wavevecspec}, and inverting the Fourier transform, we find
\begin{equation}
    \begin{split}
        \langle \delta T(t, \mathbf{r})\delta T(0, 0) \rangle &= \int \frac{\mathrm{d}^3 k}{(2\pi)^3}\;\mathrm{d}f \;\Phi(k) e^{i[\mathbf{k}\cdot (\mathbf{r}+\mathbf{U}t)-2\pi ft]},\\
        &= \int \frac{\mathrm{d}^3 k}{(2\pi)^2}\; \Phi(k) e^{i\mathbf{k}\cdot \mathbf{r}}\delta(2\pi f-\mathbf{k}\cdot\mathbf{U}).
    \end{split}
    \label{eq:tempstructure}
\end{equation}
Here, the delta function enforces Taylor's frozen turbulence hypothesis by fixing the $k$ wavevectors in the direction of the wind, setting $k_x = 2\pi f/U$ when solving the integral over $k_x$.

The remaining integrals over $\mathbf{r}$ will Fourier transform the spatial Green's functions to $k$-space. 
As we assume the temperature perturbations to be purely advected along the $x$-axis, parallel to the surface of the Earth, $G(\mathbf{r})$ is then
\begin{equation}
    G(\mathbf{r}, z_0) = -\partial_z\frac{\Theta(z-z_0)}{\sqrt{x^2+y^2+z^2}},
\end{equation}
where in Cartesian coordinates $\mathbf{r}=(x, y, z)$, and $z_0$ is the depth of the detector. Since we consider AIs sensitive to vertical accelerations in the gravitational field, we take the $z$ derivative of the function. Taking the Fourier transform to convert to wavevector space, the resulting gravitational propagator is~\cite{Brundu:2022mac}
\begin{equation}
\begin{split}
    G_\mathbf{k}(z_0) &=\\ 
    &\left[\sqrt{k_x^2+k_y^2}-ik_z\right]^{-1}\frac{2\pi i k_z}{\sqrt{k_x^2+k_y^2}}e^{-\left(\sqrt{k_x^2+k_y^2}-ik_z\right)z_0}.
\end{split}
    \label{eq:gravprop}
\end{equation}

Finally, substituting Eq.~\eqref{eq:tempstructure} and Eq.~\eqref{eq:gravprop} in Eq.~\eqref{eq:sg1} allows us to compute the power spectral density of temperature GGN
\begin{equation}
    S_g(f) = \beta^2\int \frac{\mathrm{d}^3k}{(2\pi)^2}\;\Phi(k)|G_\mathbf{k}(z_0)|^2\delta(2\pi f-\mathbf{k}\cdot\mathbf{U}).
    \label{eq:sg2}
\end{equation}

\subsubsection{Taylor’s frozen turbulence hypothesis}

Taylor's frozen turbulence hypothesis requires that the eddy turnover time, $\tau$, should be much longer than the time the eddy is seen by the detector, which primarily depends on the size of the eddy and the advecting wind speed.\footnote{Alternatives to Taylor's frozen turbulence hypothesis have also been proposed~\cite{Cheng_2017}, but we generally expect it to be valid for the frequencies considered in this work.}
Two approaches have been used to estimate the eddy turnover time.
Firstly, using dimensional arguments, it was argued that $\tau = \ell^{2/3}/\epsilon^{1/3}$~\cite{Tatarskii_1971}, where $\ell = 2\pi/k$ is the length scale of the eddy and $\epsilon$ is the rate of energy dissipation from larger to smaller turbulent structures. 
A second approach suggests $\tau = \ell/2\pi U$~\cite{Munch_1958, Wheelon_2007} for larger eddies, where $U$ is the mean wind speed.

Under the frozen turbulence hypothesis the length scale along the wind direction is given as $k^{-1} = \ell/2\pi = \omega^{-1}U$, where $\omega=2\pi f$ is the angular frequency,  effectively making this time scale $\tau = \omega^{-1}$.  For the frequency range \SIrange{0.01}{1}{Hz}, the inverse angular frequency corresponds to $\omega^{-1}\approx$~\SIrange{0.16}{16}{s}. Experimental results have concluded that eddies generally evolve on a timescale of $\tau\approx\SI{15}{\second}$~\cite{Short_2003}, which aligns well with the range of frequencies we consider. In general we expect lower frequency noise to correspond to the frozen hypothesis given that eddy evolution occurs on a slower time scale. In~\cite{Brundu:2022mac}, the condition for a laser interferometer employing the frozen turbulence hypothesis is
\begin{equation}
    \tau \gg \mathrm{max}\left(\frac{z_0}{U}, \frac{1}{\omega}\right),
    \label{eq:frozenlimit}
\end{equation}
comparing the evolutionary timescale of the eddies to the observation time a detector at depth $z_0$ sees an advected temperature perturbation, and the inverse angular frequency of the noise. 

The introduction of the detector depth factor makes for an interesting addition, especially when considering vertical atom interferometers with one atom source placed at a greater depth than the other. The noise spectrum will be dominated by the interferometer closest to the surface, which for a depth $z_0 \sim \mathcal{O}\left(\SI{10}{\metre}\right)$ and wind speed $U \sim \mathcal{O}\left(\SI{10}{\metre\per\second}\right)$, would suggest Taylor's hypothesis is generally valid on timescales of $\tau\sim \mathcal{O}\left(\SI{10}{\second}\right)$. When both atom sources are placed further underground, the modeling becomes more complex as the atoms will observe the noise on longer timescales. However, depth also mitigates the noise so that the precise modeling will become less important. 
For simplicity, we use Taylor's frozen turbulence hypothesis in our analysis and leave evaluation of its validity to future work.

\subsubsection{Temperature wavenumber spectrum}

\begin{figure}[!t]
    \centering
    \includegraphics[width=0.8\columnwidth]{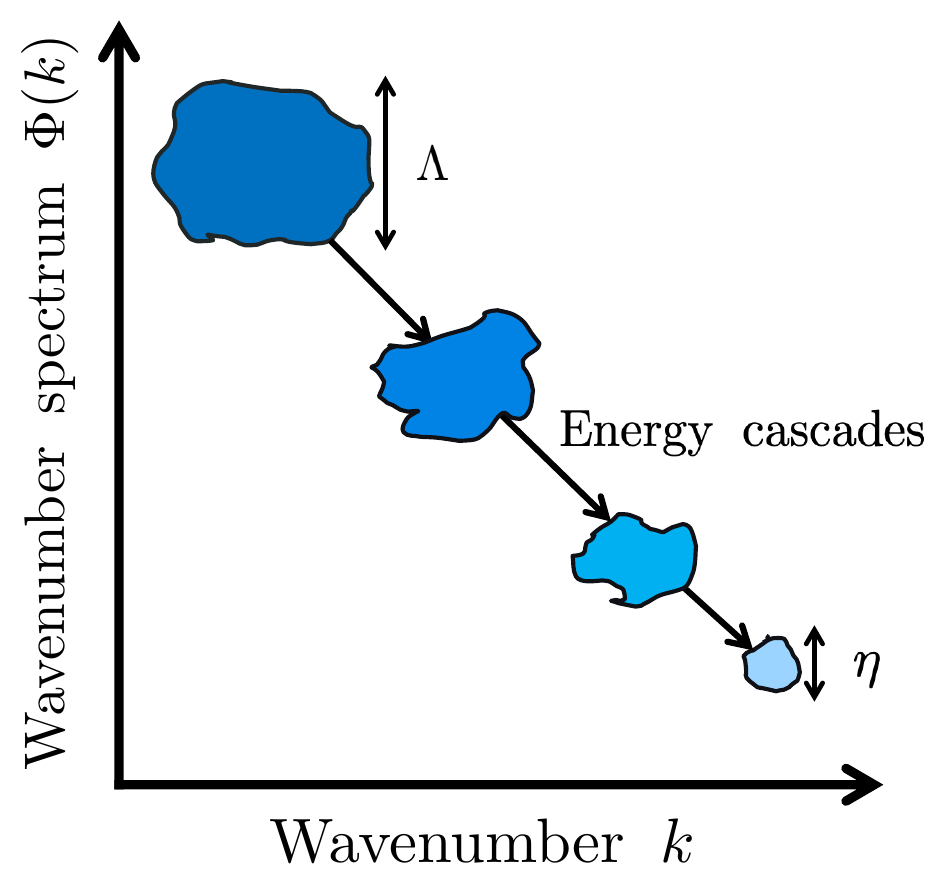}
    \caption{A wavenumber spectrum showing how energy cascades from large turbulent eddies to smaller structures in the inertial subrange. They form at the outer Kolmogorov scale $\Lambda$ and dissipate at scales smaller than $\eta$.}
    \label{fig:cascade}
\end{figure}

What remains is to determine the form of the temperature wavenumber spectrum $\Phi(k)$. Kolmogorov first described turbulence as a cascade of eddies on a wavenumber spectrum, the so-called inertial subrange~\cite{Kolmogorov_1991}. Fig.~\ref{fig:cascade} shows a representation of this. At an outer scale, large eddies form and rapidly divide into smaller eddies at larger wavenumbers until an inner scale where the smallest eddies dissipate due to atmospheric viscosity.
This model of turbulent structures forming can be very useful as the wavenumber spectrum $\Phi(k)$ can be described as a simple power law
\begin{equation}
    \Phi(k) \sim k^{-11/3},
\end{equation}
where the wavenumber $k$ is well between the outer ($\Lambda$) and inner ($\eta$) Kolmogorov scales $\Lambda^{-1}\ll k\ll\eta^{-1}$. Previous investigations into temperature GGN for laser interferometers have treated the turbulence as being entirely within the inertial subrange to simplify calculations of the noise spectrum~\cite{Creighton:2008,Brundu:2022mac}. However, the treatment is only valid down to a frequency $\sim\SI{0.15}{\hertz}$ precisely in the expected range of sensitivity for AI experiments. Thus, a model of the turbulence spectrum is required that more accurately describes the low frequency behavior.

To improve the modeling of the turbulence spectrum we examine several established models. In the inertial subrange (IS), Kolmogorov modeled the structure of turbulence as a power law, where the wavevector spectrum takes the form~\cite{Kolmogorov_1991, Greenwood:08}
\begin{equation}
    \Phi_\mathrm{IS}(k) = \frac{5}{18\pi \Gamma(1/3)} c_T^2 \,k^{-11/3},
\end{equation}
assuming $2\pi/\eta > k > 2\pi/\Lambda$. Here $c_T = \SI{0.2}{\kelvin^2\metre^{-2/3}}$ is the temperature structure constant, and $\Gamma(x)$ is the gamma function. However, the transition from the inertial subrange to lower frequency regions of parameter space will no longer follow this neat power law behavior. The frequency at which this transition occurs varies, but for typical parameters, it is on the order of $\SI{0.1}{\hertz}$, which lies directly in the sensitivity band of long baseline AI experiments. Therefore, a different model of low frequency turbulence is required to accurately model atmospheric temperature GGN for these experiments. 

The von K\'arm\'an (VK) spectrum extends the Kolmogorov subrange to lower frequencies while preserving the power law behavior at higher frequencies~\cite{Karman_1948}. It takes the form
\begin{equation}
    \Phi_\mathrm{VK}(k) = \frac{5}{18\pi \Gamma(1/3)} c_T^2\,\left(k^2+\frac{1}{\Lambda^2}\right)^{-11/6}.
\end{equation}
However, more recent studies have noted how the VK spectrum does not match up to measurements of large turbulent structure formation. The Greenwood-Tarazano (GT) model is an alternative, modeled empirically with the specific intention of improving large length scale, and thus low frequency, turbulence modeling. The full GT wavevector spectrum takes the form~\cite{Greenwood:08}
\begin{equation}
    \Phi_\mathrm{GT}(k) = \frac{5}{18\pi \Gamma(1/3)} c_T^2\, \left(k^2+\frac{k}{\Lambda}\right)^{-11/6}.
    \label{eq:GT}
\end{equation}

\begin{figure}[!t]
    \centering
    \includegraphics[width=\columnwidth]{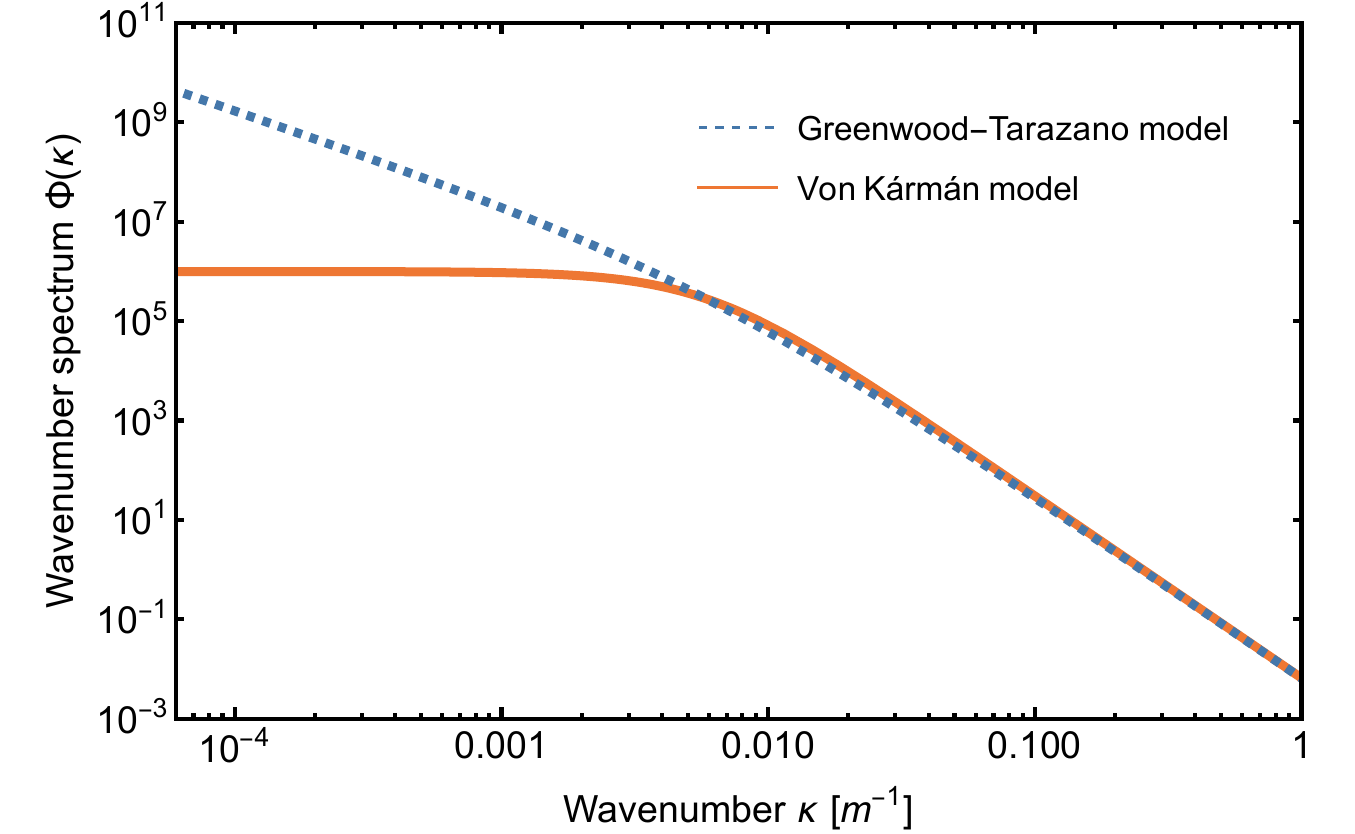}
    \caption{Comparison of the Greenwood-Tarazano model to the von K\'arm\'an model of temperature structure. Both are normalized wavenumber spectra showing how they have identical scaling behavior in the inertial subrange but differ at large scales (low frequencies).}
    \label{fig:GT_VK_comp}
\end{figure}

A comparison between the wavenumber spectrum for the GT and VK models is shown in Fig.~\ref{fig:GT_VK_comp}.
At higher wavenumbers, and thus higher frequencies, both spectra collapse to the expected power law of the inertial subrange. However, while the VK spectrum tends to a constant value at low frequency, the GT spectrum continues to grow at a slower rate, which more accurately aligns with empirical data~\cite{Greenwood:08, Wheelon_2007}. 

Each spectrum relies on the outer scale of turbulence $\Lambda$. The definition of the outer scale has been widely debated and varies depending on specific environmental conditions. In general, it scales with height $z$~\cite{Brundu:2022mac}. However, at significant heights above the surface the relationship between the outer scale, height, and atmospheric stability changes~\cite{Tofsted_2000}. Small values of $\Lambda$ restrict the size of turbulent structures that can form in the atmosphere, limiting the impact of noise. Beyond a few tens of meters, the outer scale has a negligible impact on the temperature noise spectrum. As we consider the most significant contribution to the noise coming from the surface boundary layer, we approximate the outer scale as the Obukhov length. We will further justify this choice in Sec.~\ref{sec:atmoggn-results}.

\subsubsection{Strain noise spectrum}

Substituting $\Phi_\mathrm{GT}(k)$ into Eq.~\eqref{eq:sg2} gives us the full equation to evaluate. The delta function enforces Taylor's frozen turbulence hypothesis for the wavenumber in the direction of the wind. The integral can then be solved by numerically integrating the expression over $k_y$ and $k_z$. This expression gives the spectrum of gravity perturbations at a particular measurement point. To convert it to the spectrum of noise in a differential experiment, we consider the locations of our test masses relative to each other. In horizontal long-baseline laser interferometer experiments such as LIGO, GGN is typically uncorrelated at each test mass site. This is due to the test masses being sufficiently separated on the surface of the planet as well as having a peak sensitivity in a higher frequency band than atom interferometers. The noise in this uncorrelated case simply adds linearly with an independent contribution from each test mass $k$. The strain noise for the uncorrelated noise case is then
\begin{equation}
    S_h^\mathrm{uncor.}(f) = \frac{1}{\omega^4 L^2}\sum_k S_g(f),
\end{equation}
where $\omega = 2\pi f$ is the angular frequency of interest and $L$ is the baseline length separating the test masses.

However, in a vertical AI experiment, we expect a source of horizontally propagating noise to be correlated along the baseline. The atom sources each feel the effects of the same atmospheric turbulence above them at different depths. Thus, we subtract the noise to find the correlated noise spectrum. The strain noise for the correlated noise case is
\begin{equation}
    S_h^\mathrm{corr.}(f) = \frac{1}{\omega^4 L^2}\left[S_g(f, z_0) - S_g(f, z_0+\Delta z)\right],
    \label{eq:corrSh}
\end{equation}
where the gravitational noise spectrum is a function of frequency and depth with $z_0$ being the minimum distance between the uppermost atoms to the surface and $\Delta z$ being the separation between the atom sources.\footnote{We note the atoms travel vertically during the sequence but this motion is accounted for in our calculations.}

\section{Impact of atmospheric GGN on atom interferometers}
\label{sec:atmoggn-results}

Having established the mathematical framework for how atmospheric pressure waves and temperature fluctuations generate GGN through density perturbations, and derived the corresponding strain noise spectrum for correlated noise sources, we now turn our attention to characterizing the impact on a vertical AI configuration.

\subsection{From GGN to phase}

Noise can enter the AI's phase shift through two primary mechanisms: direct imprinting of laser phase onto the atoms during atom-light interactions, and from gradients in the gravitational or magnetic background fields during the free propagation of the atoms that arises from the action over the classical trajectory of the atom cloud~\cite{Hogan:2008}. The transfer function, $T_\phi$, captures these effects and converts between the strain spectrum and phase spectrum 
\begin{equation}
    S_\phi (f) =  T_\phi^2  S_h(f).
 \label{eq:Sphi}
\end{equation}
The transfer function is sequence dependent and for the experiments we consider in this work is
\begin{equation}
    T_\phi =  2 n \kappa L \cos\left[\omega T\right] \sin^2\left[\frac{\omega T}{2}\right],
    \label{eq:transfer}
\end{equation}
where $n$ is the order of momentum transfer between the AI arms, $\kappa$ is the laser wavevector, $L$ is the baseline length of the interferometer and $T$ is the free propagation time between the beam splitter and mirror pulses. 

Here we assume a broadband three pulse Mach-Zehnder sequence that is enhanced by a large momentum transfer (LMT) factor $n$ and neglect the finite speed of light traveling between the interferometers. This serves as a good approximation for a real $n$-pulse sequence. The formalism established here makes finding the atmospheric pressure and temperature noise present in an arbitrary interferometry sequence straightforward. Future work may also consider the noise as seen in other types of sequences, for example resonant as opposed to broadband. For more details on these alternative sequences, see, e.g.~\cite{Wang:2024puy}. 

The origin of the $\kappa$ term in Eq.~\eqref{eq:transfer} requires careful explanation, as it depends on the specific atom-light interactions used in the AI sequence. 
In this paper, we focus on single-photon (or more generally single-direction) atom-light interactions where $\kappa = \omega_a /c$, with $\omega_a$ representing the atomic transition frequency~\cite{Yu:2010ss,Graham:2013,Carman:2024}.
The total phase shift of the AI is directly sensitive to the laser phase imprinted on the atoms at each interaction and to the difference between the laser frequency and the atomic frequency.\footnote{When  on resonance, the detuning is zero, and $\omega_{\mathrm{laser}} = \omega_a$.} 
Fluctuations in the laser phase arise from two sources: laser instability and vibrations of the optics between the laser and the atoms.
However, in a gradiometer configuration, these noise sources become common to both AIs as the light propagates down the baseline. 
Therefore, the differential measurement effectively suppresses this common-mode noise~\cite{Rudolph:2020,Hu:2019}.
Further details on atom-light interactions in interferometry can be found in~\cite{Kasevich:1991,Giltner:1995,Muller:2009,Altin:2013,Hu:2019,Rudolph:2020}.

Using the infrasound generated gravitational potential, Eq.~\eqref{eq:6}, we calculate the AI phase shift implementing a semi-classical perturbative calculation~\cite{Hogan:2008}
\begin{align}
    \delta \phi_{\mathrm{infra}}(z_0) = {}&\frac{1}{\omega^2L}\Bigg[\frac{8\pi\alpha}{k^2}  \qty(k_\varrho - 2k_z \Theta(z_0) e^{k_\varrho z_0}\sin(k_z z_0))\nonumber\\
    &\quad \times e^{-k_\varrho \abs{z_0}} \delta p(\omega)\Bigg] T_\phi.
   \label{eq:8}
\end{align}
We make a conservative assumption here such that there is no minimum distance between the density fluctuations and the atom cloud ignoring any effects of infrasound screening from exterior structures or the vacuum system supporting the trajectory of the atoms. There can also be a lower frequency cut-off arising from the finite size of the atmosphere that we do not account for here.

The infrasound GGN spectrum in an atom gradiometer is then given by
\begin{equation}
    S_{\phi,\rm infra} = \delta\phi_{\rm infra}^2(z_0)-\delta\phi_{\rm infra}^2(z_0+\Delta z),
\end{equation}
taking the difference of this spectra for our two interferometers gives the gradiometer infrasound noise spectrum.
For temperature noise, we simply convert the strain spectra derived in Eq.~\eqref{eq:corrSh} using the transfer function as in Eq.~\eqref{eq:Sphi}.

\begin{table}[!t]
    \begin{tabular}{c c c c c c c} 
    \toprule
    Isotope & $\kappa$ [m$^{-1}$] & $L$ [m] & $T$ [s] & $n$ & $z_0$ [m] & $\delta\phi$ [rad/$\sqrt{\mathrm{Hz}}$] \\ 
    \midrule
    ${}^{87}$Sr & $9.002\times 10^{6}$ & $10^2$ & 1.4 & $10^3$ & 10 & $10^{-4}$ \\ 
    \hline
    ${}^{87}$Sr & $9.002\times 10^{6}$ & $10^3$ & 1.4 & $10^3$ & 10 & $10^{-5}$ \\ 
    \bottomrule
    \end{tabular}
    \caption{Design parameter characteristic of two long-baseline atom interferometer experiments. Isotope denotes the species of atom used; $\kappa$ is the effective wavevector of the laser; $L$ is the baseline length separating two atom sources; $T$ is the interrogation time between interferometer pulses; $n$ is the number of large-momentum transfer pulses; $z_0$ is the minimum distance between the upper interferometer and noise source; $\delta\phi$ is the expected level of atom shot noise in the experiment.}
    \label{tab:params}
\end{table}

In the following subsections, we use design parameters proposed for upcoming terrestrial experiments with proposed baselines of $\SI{100}{\metre}$ and $\SI{1}{\kilo\metre}$. The complete list of parameters is given in Table~\ref{tab:params}.

\begin{figure*}[ht]
    \begin{minipage}{0.49\textwidth}
    \centering
    \includegraphics[width=\textwidth]{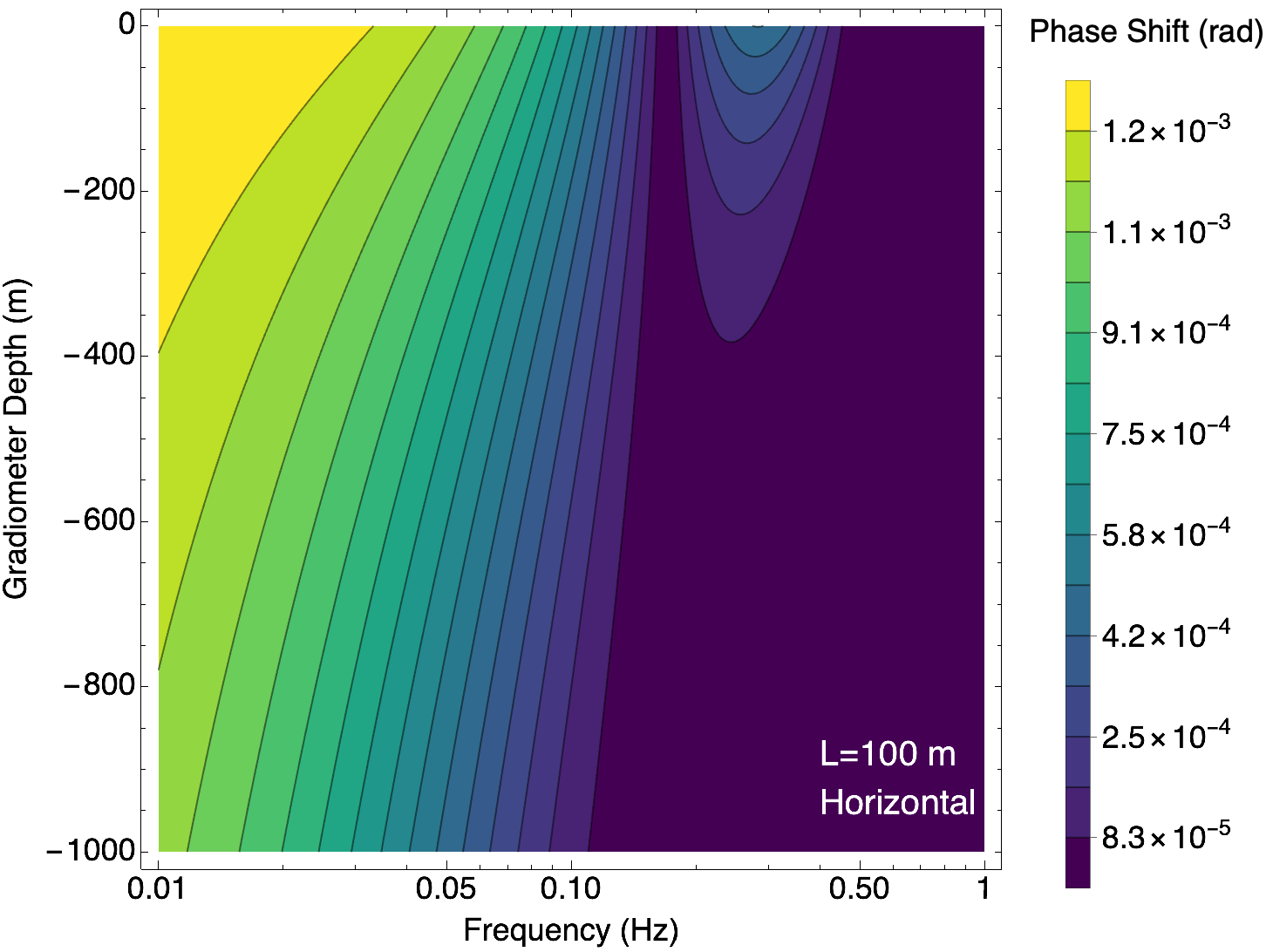}
    \end{minipage}%
    \hfill
    \begin{minipage}{0.49\textwidth}
    \centering
    \includegraphics[width=\linewidth]{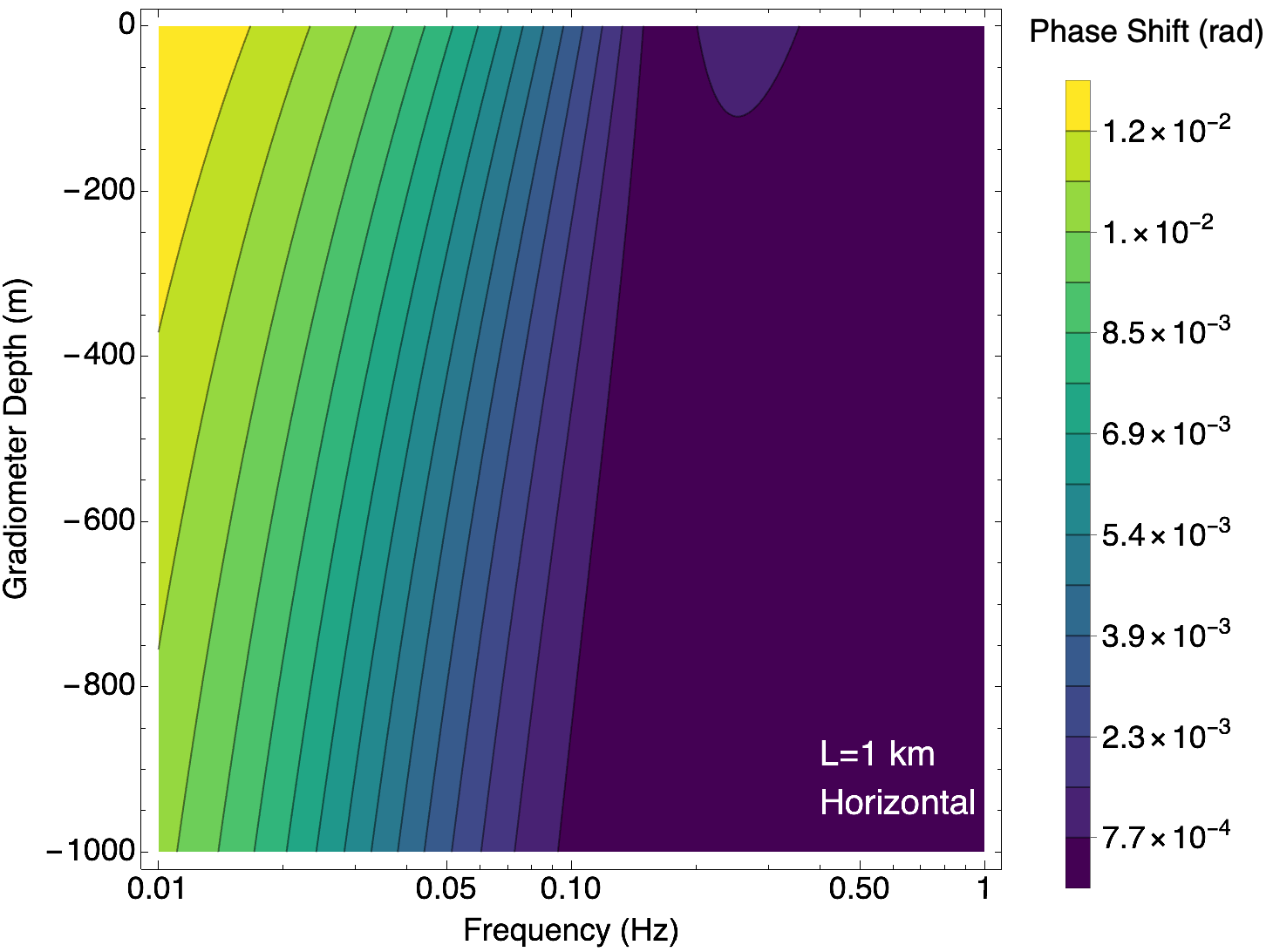}
    \end{minipage}\\
    \begin{minipage}{0.49\textwidth}
    \centering
    \includegraphics[width=\linewidth]{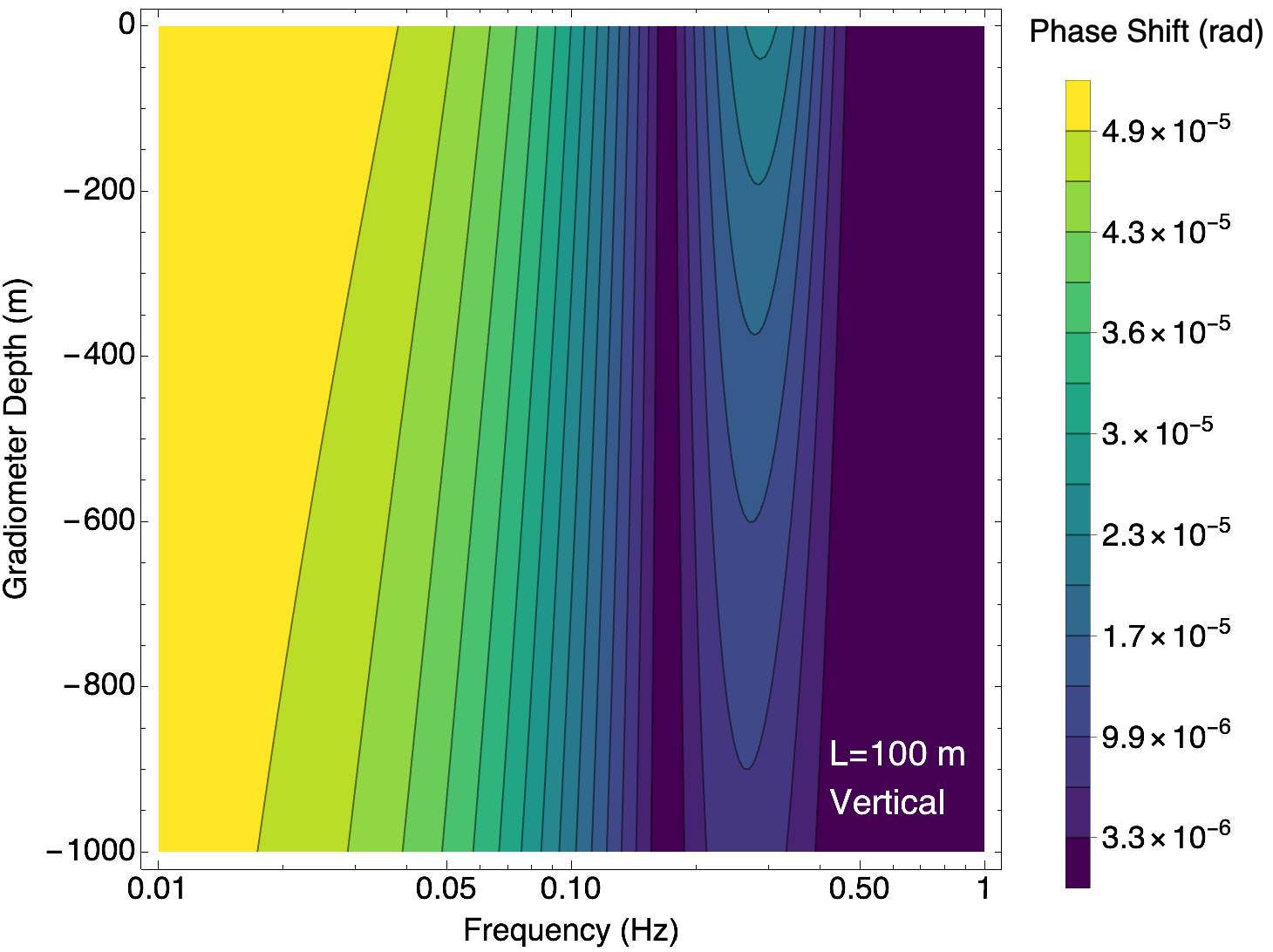}
    \end{minipage}%
    \hfill
    \begin{minipage}{0.49\textwidth}
    \centering
    \includegraphics[width=\linewidth]{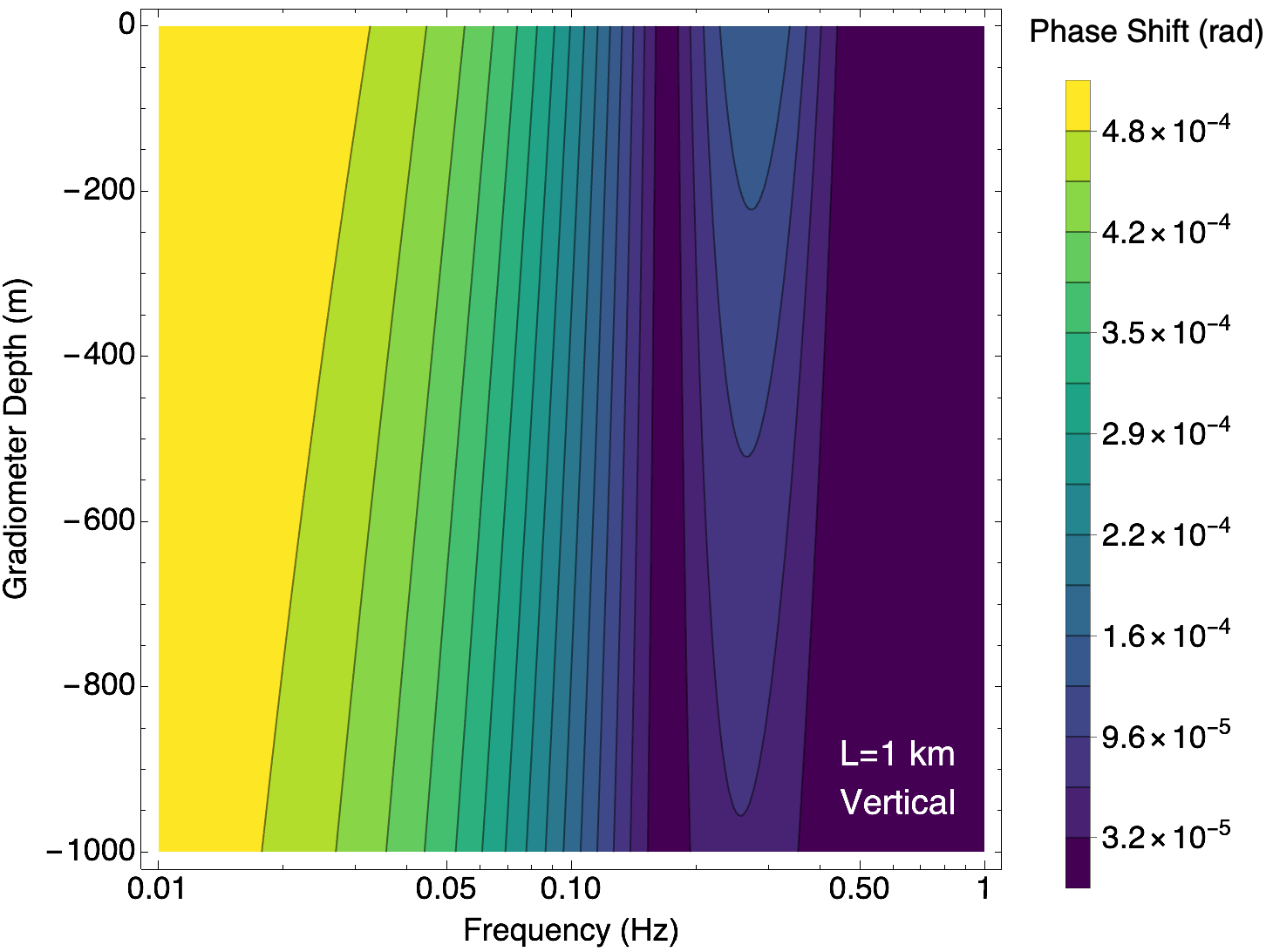}
    \end{minipage}\\
    \begin{minipage}{0.49\textwidth}
    \centering
    \includegraphics[width=\linewidth]{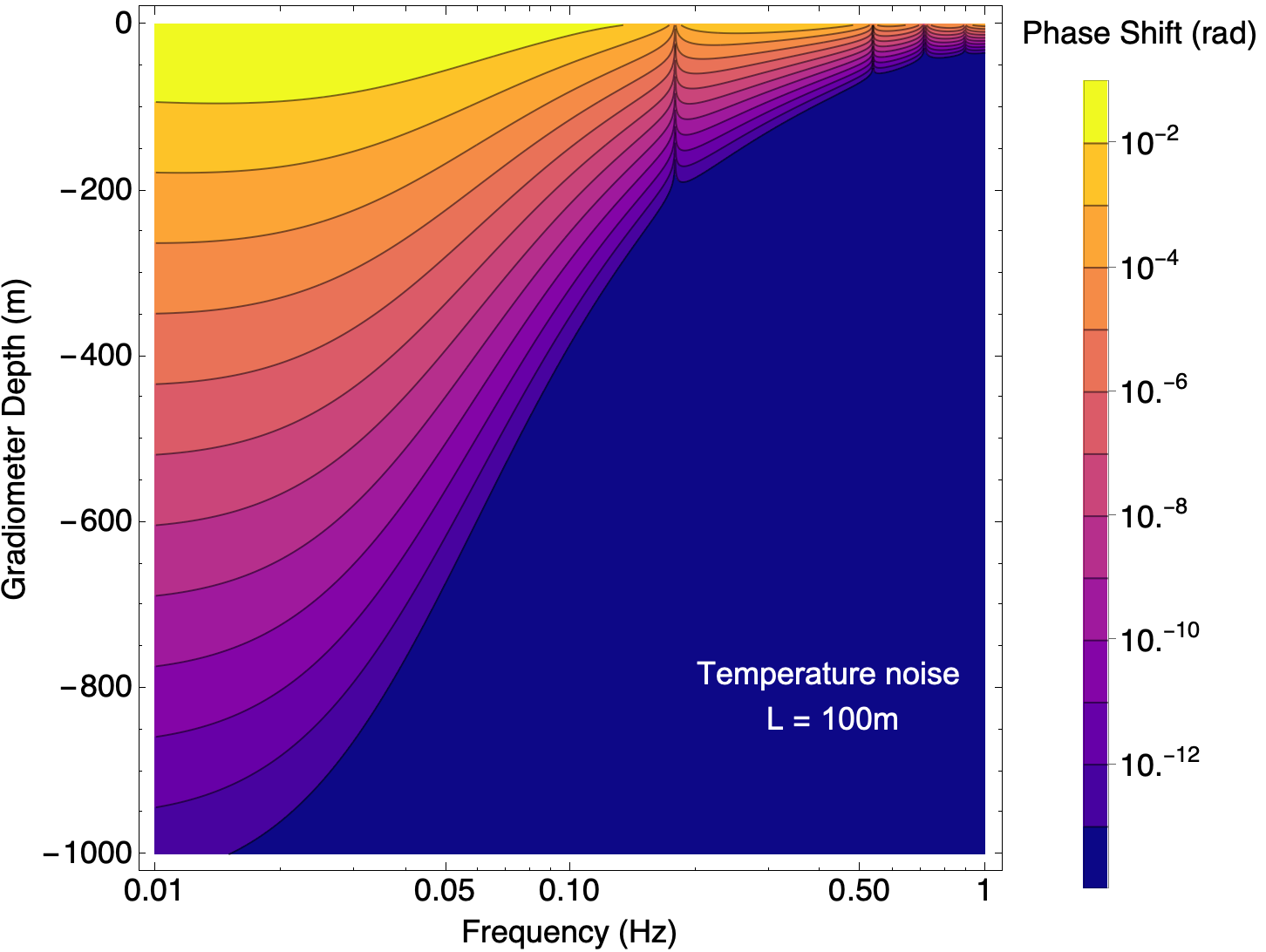}
    \end{minipage}
    \hfill
    \begin{minipage}{0.49\textwidth}
    \centering
    \includegraphics[width=\linewidth]{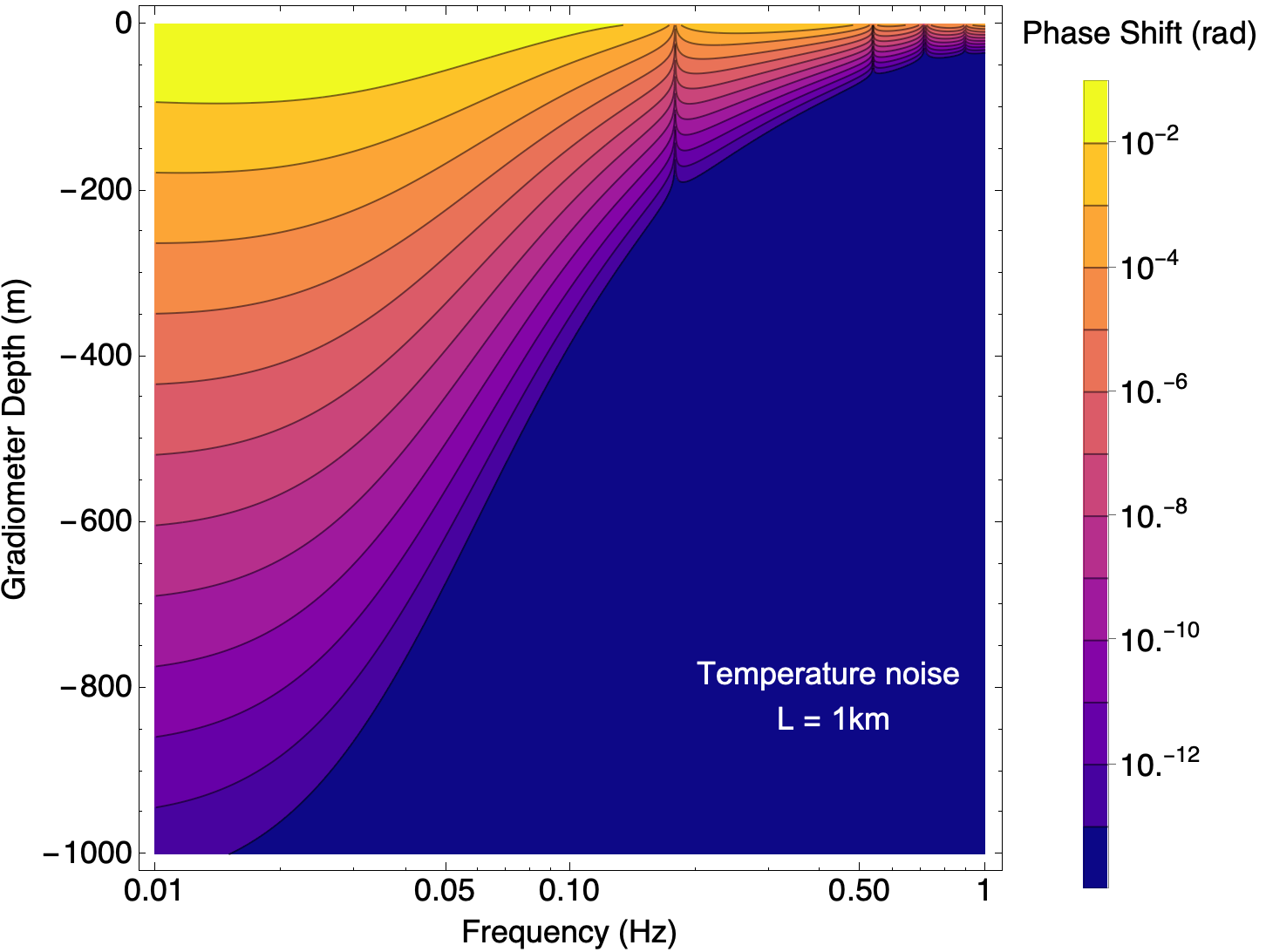}
    \end{minipage}\hfill
    \caption{Atom interferometer gradiometer phase shift contours for gradiometer depth versus frequency. The left column uses experimental parameters for the first row of Table~\ref{tab:params} while the right column uses experimental parameters from the second row of Table~\ref{tab:params}. In either case phase shifts larger than the atom shot noise sensitivity goals of \SI{1e-4}{rad} and \SI{1e-5}{rad} (see last column of Table~\ref{tab:params}) would directly limit our science reach. The depth is defined as distance of interferometer closest to the surface.  Top row: plots are the contours for pressure induced phase shift from a nearly horizontal incident pressure wave, $\theta = 7\pi/16$. Middle row: plots are the contours for pressure induced phase shift from a nearly vertical incident pressure wave, $\theta = \pi/16$. Bottom row: plots are for temperature induced phase shift, we assume typical atmospheric parameter values:  average air density $\rho_0 = \SI{1.3}{\kilo\gram\per\metre\cubed}$, average temperature $T_0 = \SI{300}{\kelvin}$, outer scale $\Lambda = \SI{170}{\metre}$, and average wind speed $U = \SI{20}{\metre\per\second}$.}
    \label{fig:depth_gradphaseshift}
\end{figure*}

\begin{figure*}[ht]
    \begin{minipage}{0.49\textwidth}
    \centering
    \includegraphics[width=\textwidth]{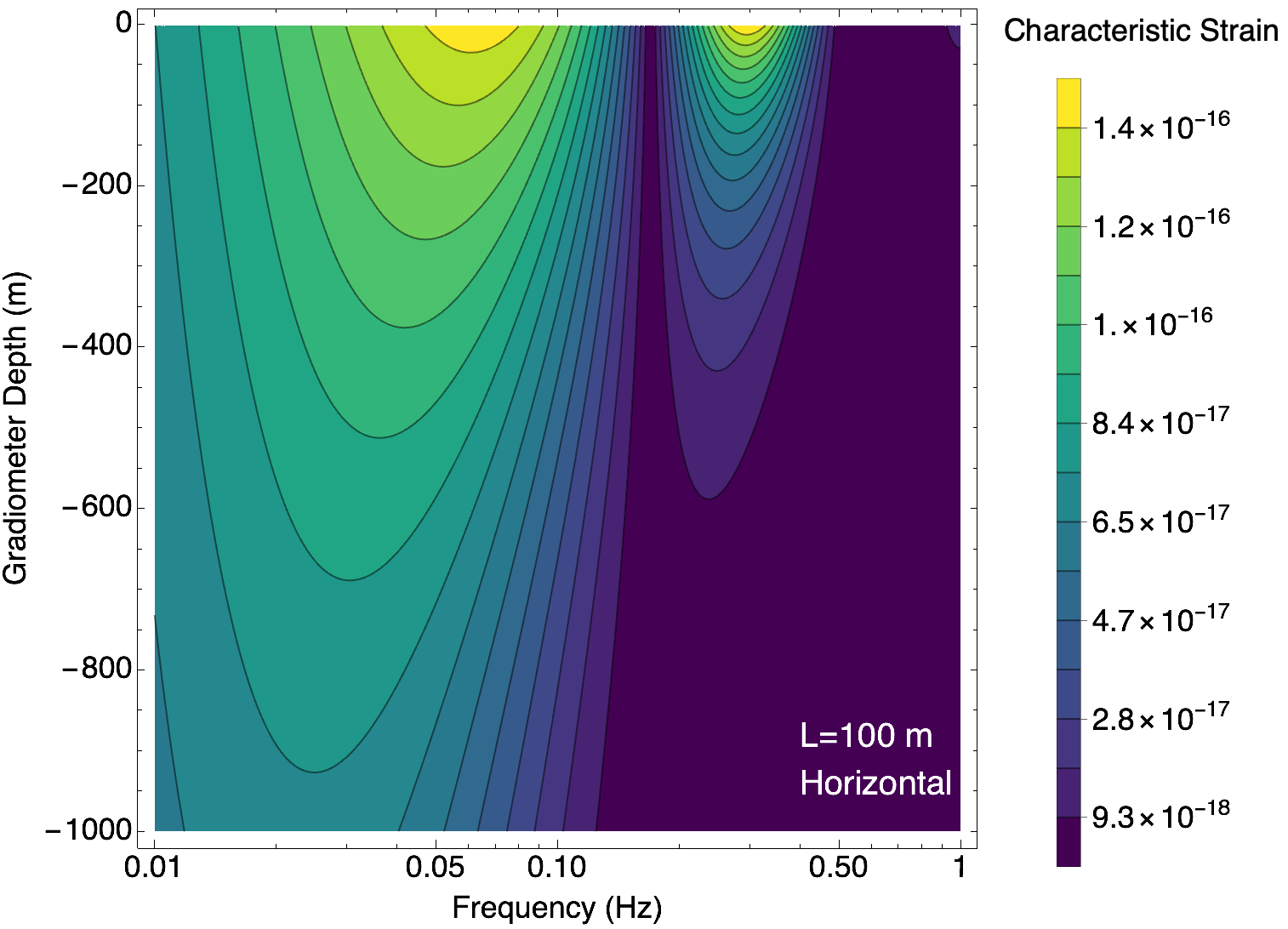}
    \end{minipage}%
    \hfill
    \begin{minipage}{0.49\textwidth}
    \centering
    \includegraphics[width=\linewidth]{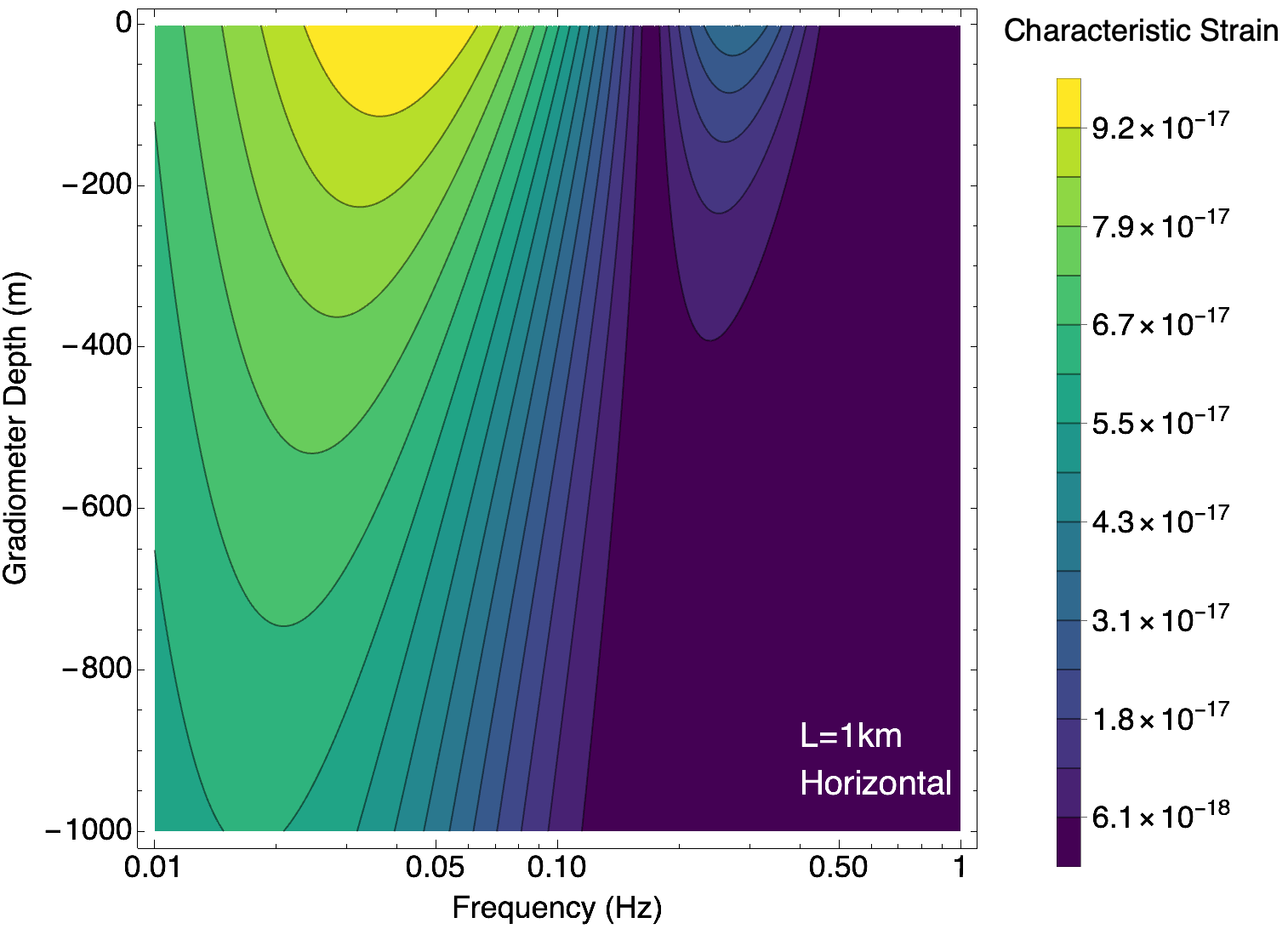}
    \end{minipage}\\
    \begin{minipage}{0.49\textwidth}
    \centering
    \includegraphics[width=\textwidth]{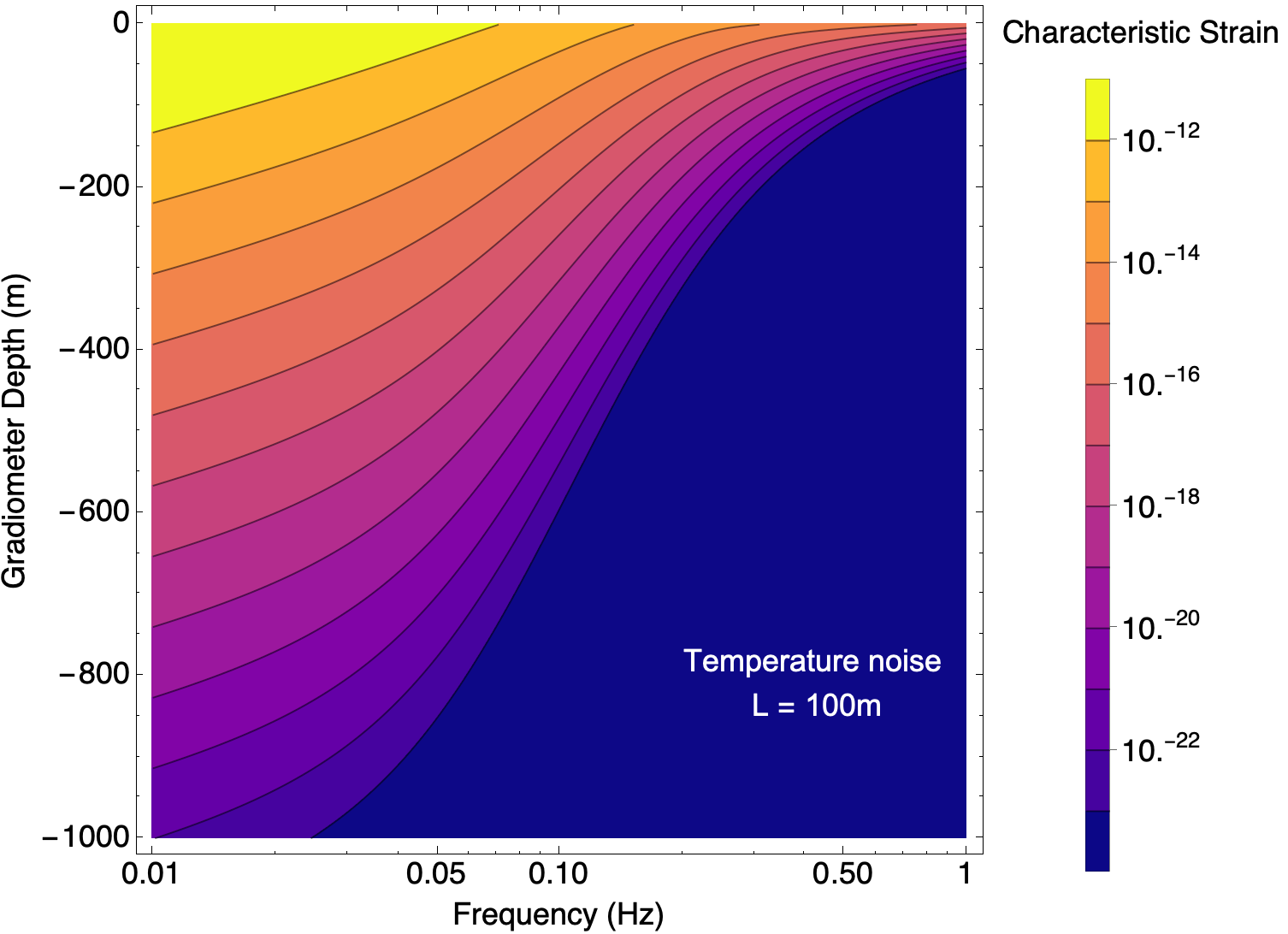}
    \end{minipage}%
    \hfill
    \begin{minipage}{0.49\textwidth}
    \centering
    \includegraphics[width=\linewidth]{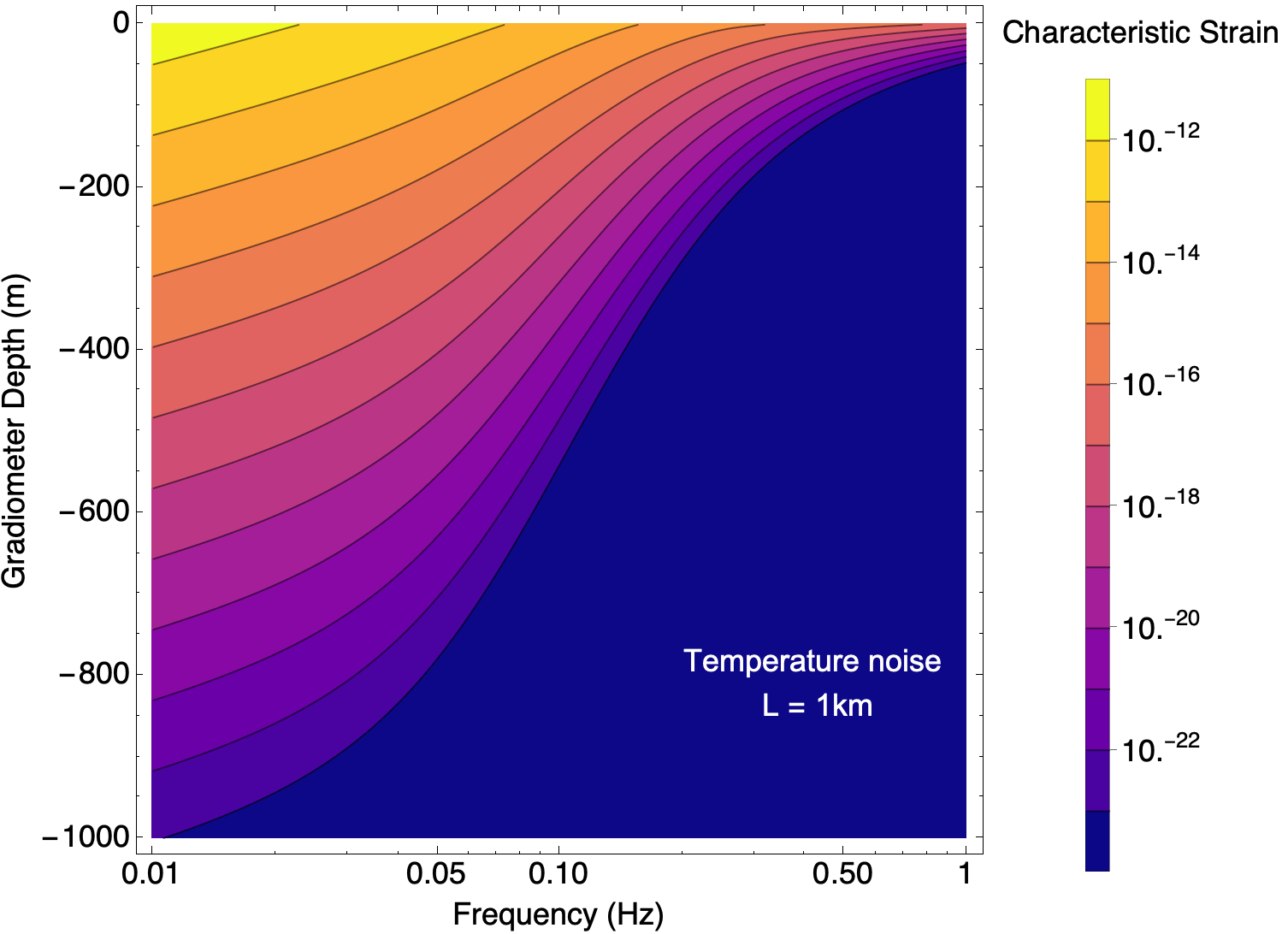}
    \end{minipage}
    \caption{Characteristic strain contours for gradiometer depth versus frequency sourced by horizontally propagating pressure plane waves ($\theta = 7\pi / 16$) and temperature fluctuations assuming the same experimental parameters as in figure~\ref{fig:depth_gradphaseshift} with the phase shift converted into characteristic strain. Again the left column uses the first row of Table~\ref{tab:params} and the right column uses the second row of Table~\ref{tab:params}. Top row: plots of pressure induced noise with contours in units of characteristic strain. Bottom row: plots of temperature induced noise with contours in units of characteristic strain. In both lower plots we assume typical atmospheric parameter values with average air density $\rho_0 = \SI{1.3}{\kilo\gram\per\metre\cubed}$, average temperature $T_0 = \SI{300}{\kelvin}$, outer scale $\Lambda = \SI{170}{\metre}$, and average wind speed $U = \SI{20}{\metre\per\second}$.}
    \label{fig:infra-gradstrain}
\end{figure*}

\subsection{Depth profile of atmospheric GGN}
\label{subsec:depth-atmoggn}

To investigate the impact of depth on the gradiometer (differential) phase shift, we explore different configurations of baseline length, depth, and infrasound incident angle, assuming typical ambient pressure and temperature conditions. 

Fig.~\ref{fig:depth_gradphaseshift} shows contour plots of the gradiometer phase as functions of the gradiometer depth and frequency, where the colors show the phase shift amplitude.
The gradiometer depth is defined as the distance of the upper AI below the surface. The upper set of four plots show the gradiometer phase from pressure induced phase noise, while the lower two plots show the phase for temperature induced phase noise. The left column assumes the parameters for the \SI{100}{m} configuration, with parameters given in Table~\ref{tab:params}, while the right column uses the parameters for \SI{1}{km} configuration.

We first discuss the noise sourced by infrasound waves, as in Eq.~\eqref{eq:6}, which is illustrated in the top and middle rows of  Fig.~\ref{fig:depth_gradphaseshift}. 
As the two AI are not co-located, they will experience different gradients and thus accumulate different phases. The phase shift for each AI is computed using Eq.~\eqref{eq:8} at two different initial heights separated by a baseline $L$ for horizontal and vertical infrasound waves with an incidence angle $\theta$ relative to the surface normal.  As in Fig.~\ref{fig:pressure-accel}, we choose a single Fourier component with a fluctuation amplitude $\delta p(\omega) = \SI{e-3}{mbar}$~\cite{Bowman:2005}. Nearly horizontal infrasound waves are obtained when $\theta = \pi/2$, and nearly vertical infrasound waves when $\theta = \pi/16$. 

While depth appears to somewhat mitigate the impact of infrasound noise, we find that the differential phase-shift structure depends most on the frequency range. The infrasound noise appears to fall to relatively insignificant levels above a frequency $\sim\SI{0.1}{\hertz}$, where the frequency band of interest for GW searches is \SIrange{0.1}{3}{\hertz}. However, a feature depending on the AI interrogation time appears around $\sim\SI{0.25}{\hertz}$. Comparing the upper four plots, we also note that this feature has a strong dependence on the propagation of the sourcing infrasound wave. For the vertically propagating infrasound wave case, the noise is maximized across the frequency range. We interpret this as the resulting gravitational perturbations penetrating further underground than for horizontally propagating waves. Vertical infrasound waves act as a standing wave at the interface of the air and ground generating longer range impacts underground, but with lower overall magnitude. Between the $L=\SI{100}{\metre}$ and $L=\SI{1}{\kilo\metre}$ experiments, the noise profiles are similar with a slight reduction in the noise for the shorter baseline case due to the lower AI being further from the source. In general, the AI closest to the surface gives the dominating contribution to the overall noise as the atoms map out steeper gravity gradients.

\begin{figure*}[!t]
    \begin{minipage}{0.49\textwidth}
    \centering
    \includegraphics[width=\linewidth]{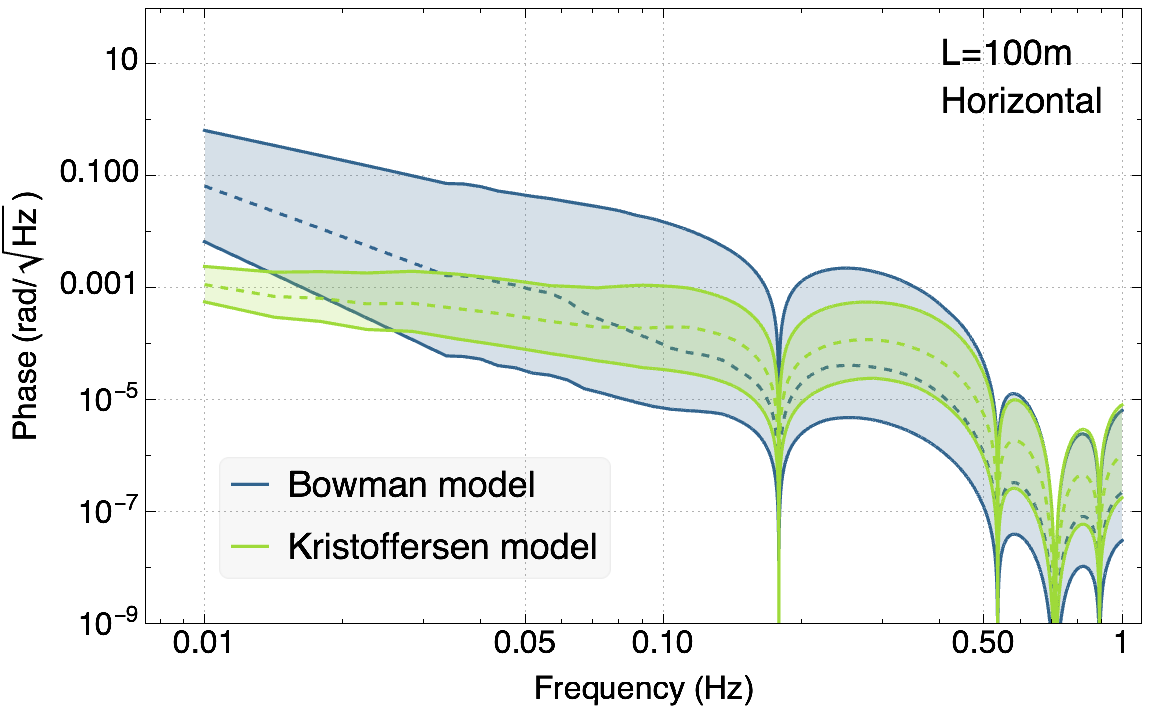}
    \end{minipage}%
    \hfill
    \begin{minipage}{0.49\textwidth}
    \centering
    \includegraphics[width=\linewidth]{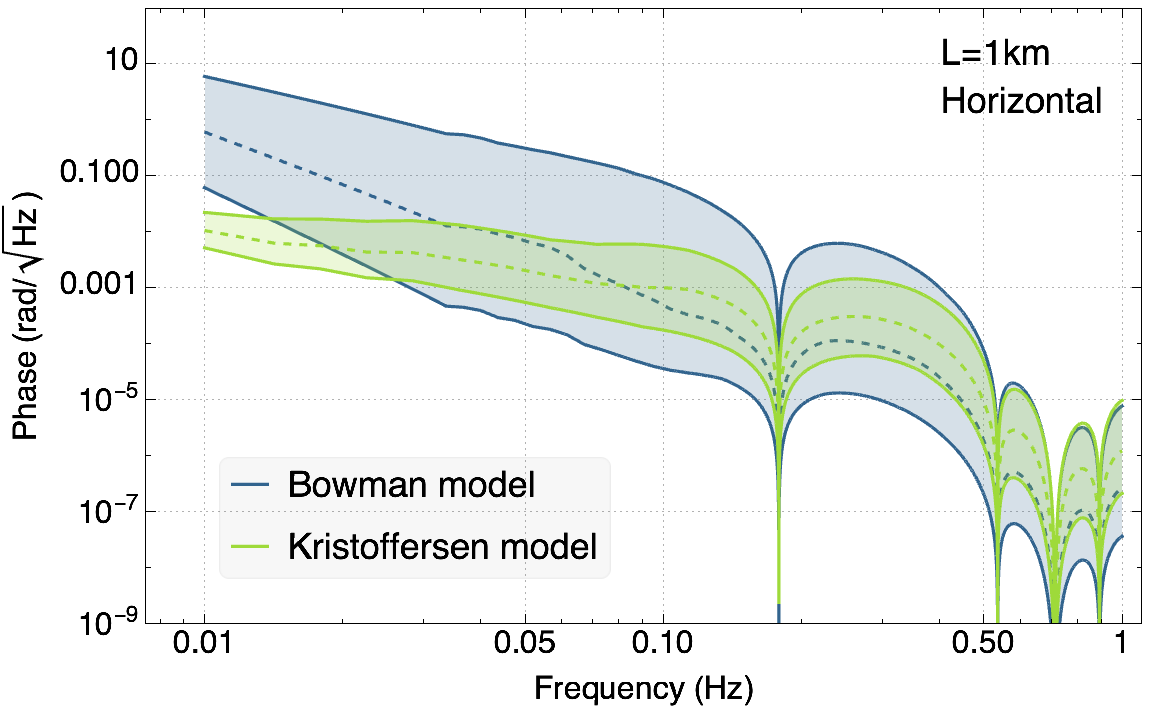}
    \end{minipage}
    \caption{Gradiometer phase amplitude spectral density for the incoherent Bowman infrasound model and the coherent Kristoffersen model. Outer curves are the upper and lower bounds of the models and dashed lines are the median values. We assume the horizontal propagation of infrasound waves as these produce a larger induced noise source ($\theta = 7\pi/16$). Left: plot using parameters from first row of Table~\ref{tab:params}. Right: plot using experimental parameters from second row of Table~\ref{tab:params}. Target phase sensitivities from atom shot noise for the \SI{100}{m} and \SI{1}{km} baselines are \SI{1e-4}{rad/\sqrt{Hz}} and \SI{1e-5}{rad/\sqrt{Hz}}, respectively.}
    \label{fig:infrasound-phaseASD}
\end{figure*}

We next discuss the noise sourced by temperature fluctuations, which is illustrated in the bottom row of Fig.~\ref{fig:depth_gradphaseshift}. 
To find the phase noise spectrum for temperature fluctuations in an atom interferometer, we numerically calculate the spectrum directly using Eq.~\eqref{eq:Sphi} and Eq.~\eqref{eq:corrSh}. Unlike pressure fluctuations, there is no general analytical solution for calculating the noise as in Eq.~\eqref{eq:7}. We assume the temperature fluctuations are advected by the wind parallel to the ground, ignoring any effects from surrounding buildings and infrastructure as with the pressure fluctuations. As the AIs are separated along a vertical baseline perpendicular to the movement of the turbulent eddies, their incident direction has no impact on the noise they induce. 

The bottom row of Fig.~\ref{fig:depth_gradphaseshift} shows temperature phase shift contour plots for vertical $\SI{100}{\metre}$ and $\SI{1}{\kilo\metre}$ baseline atom interferometer experiments using the parameters listed in Table~\ref{tab:params}. The average wind speed is a somewhat aggressive $U=\SI{20}{\metre\per\second}$ to simulate a regime in which this noise will be more problematic to the experiment, and the outer scale is $\Lambda = \SI{170}{\metre}$. Again, it is clear to see that the noise structure heavily depends on the frequency as with infrasound; however,  gradiometer depth impacts the noise level to a much greater extent. Each contour represents an order of magnitude improvement in the noise as the horizontally advected temperature perturbations have no extent vertically into the ground, unlike infrasound. The dips in noise sensitivity occurring at higher frequencies depend on the interrogation time and occur at the zeros of the transfer function defined in Eq.~\ref{eq:Sphi}. While there is a slight difference in noise between the $L=\SI{100}{\metre}$ and $L=\SI{1}{\kilo\metre}$ experiments, the AI closest to the surface again dominates. The significant fall off of noise with depth means the lower AI contributes little to the overall gradiometer noise.

For completeness, we plot the characteristic strain equivalent plots to the depth profiles of infrasound and temperature GGN. The counterpart to Fig.~\ref{fig:depth_gradphaseshift} can be seen in Fig.~\ref{fig:infra-gradstrain}. These are effectively calculated by dividing the phase spectra by the transfer function defined in Eq.~\eqref{eq:transfer}. The strain spectra are useful to examine as they present the atmospheric GGN without any detector-dependent features. To examine the noise for a different AI sequence, simply multiply these spectra by the relevant transfer function.

\subsection{Global models of infrasound noise}\label{subsec:global-infra}

For global expected limits of infrasound noise we employ two global ambient models from the atmospheric sciences literature. An international monitoring system comprised of 55 sensor stations was established over the past two decades to record ambient infrasound around the globe. For our study we refer to an older high and low noise model established by Bowman \textit{et al}.~\cite{Bowman:2005}. Their survey used 21 infrasound arrays in the frequency band \SIrange{0.03}{7}{Hz} over one year. We also investigate an updated model by Kristoffersen \textit{et al}.~\cite{Kristoffersen:2022}. Their study used 53 of the International Monitoring System sensor arrays over the frequency range \SIrange{0.01}{4}{Hz} over a period of 3-4 years. The Bowman model analyzed the ambient infrasound data with a focus on incoherent noise to establish upper and lower limits globally, whereas the Kristoffersen model explores the coherent ambient infrasound noise implementing newer array analysis techniques. Both groups make use of power spectral density (PSD) analysis methods to generate histograms and probability density functions for the sensor arrays. 

We take the pressure-amplitude spectral densities computed for the models' upper, lower, and median values as our input. With these Fourier amplitudes as $\delta p(\omega)$ in Eq.~\eqref{eq:8} we calculate the per-shot gradiometer phase shift amplitude spectral density for the parameters given in Table~\ref{tab:params}, assuming typical ambient conditions and pressure waves propagating horizontally. The results are shown in Fig.~\ref{fig:infrasound-phaseASD}.

In this format these AI global infrasound noise curves can be compared with other potential noise sources such as {\it seismic GGN}, discussed in Sec.~\ref{subsec:compare-seismic}, when presented as amplitude spectral densities. Spectral densities are also useful when modeling background noise in the presence of a signal, e.g. GW or dark matter, and comparing data analysis methods. Finally, they can be used as a criterion for site selection by taking measurements at a proposed installation site of a detector and comparing with these global limits to inform potential phase noise in the AI from incoherent and coherent infrasound.

When using global limits, such as those plotted in Fig.~\ref{fig:infrasound-phaseASD}, to judge suitability of a potential host site, the stationarity of the projected global limits for atmospheric GGN should be accounted for. The data for the incoherent ambient infrasound noise of Bowman~\cite{Bowman:2005} is from January 2003 -- January 2004 and the coherent data of Kristoffersen~\cite{Kristoffersen:2022} includes data from 2001 --2020. In these analyses, it is shown that the ambient amplitude spectral densities have strong seasonal dependence, where amplitudes can vary by up to two orders of magnitude. These considerations together support the need for continuous local and regional atmospheric monitoring during an experimental search campaign, and of long duration site measurements when utilizing the global limits for site feasibility in order to capture the true variance of the noise levels.
This conclusion will again be highlighted in Sec.~\ref{subsec:compare-seismic}. 

\subsection{Temperature noise using ERA5 data}\label{subsec:site-compare}

\begin{figure*}[ht]
    \centering
    \includegraphics[width=0.32\textwidth]{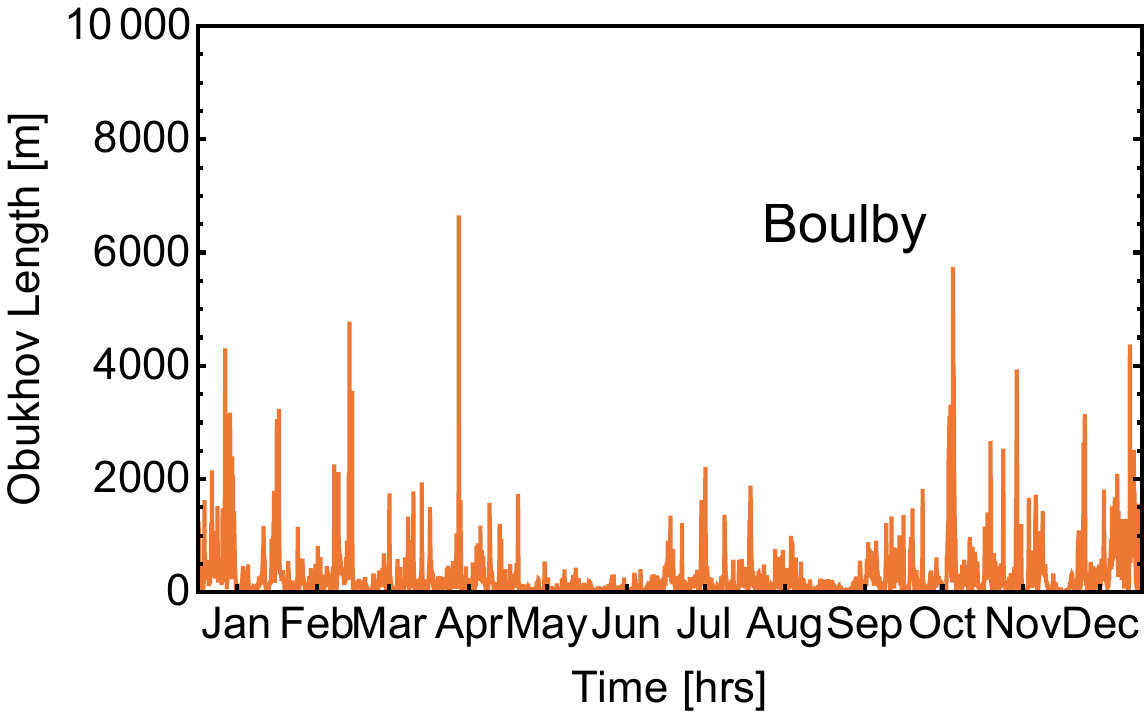}
    \includegraphics[width=0.32\textwidth]{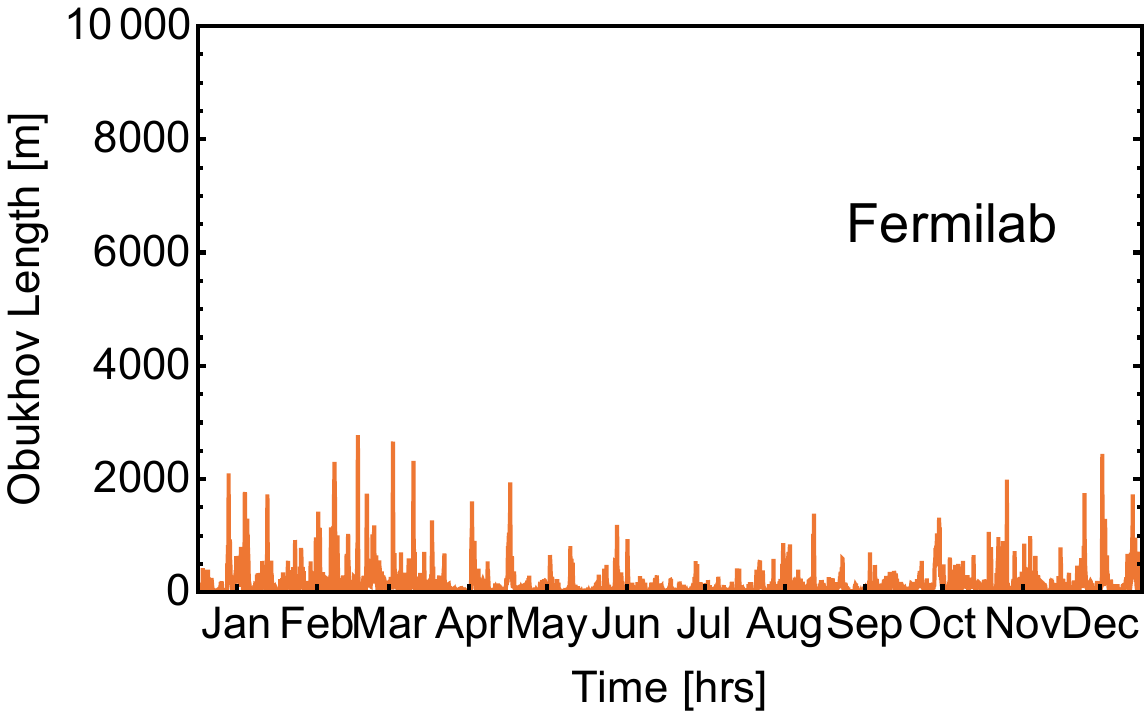}
    \includegraphics[width=0.32\textwidth]{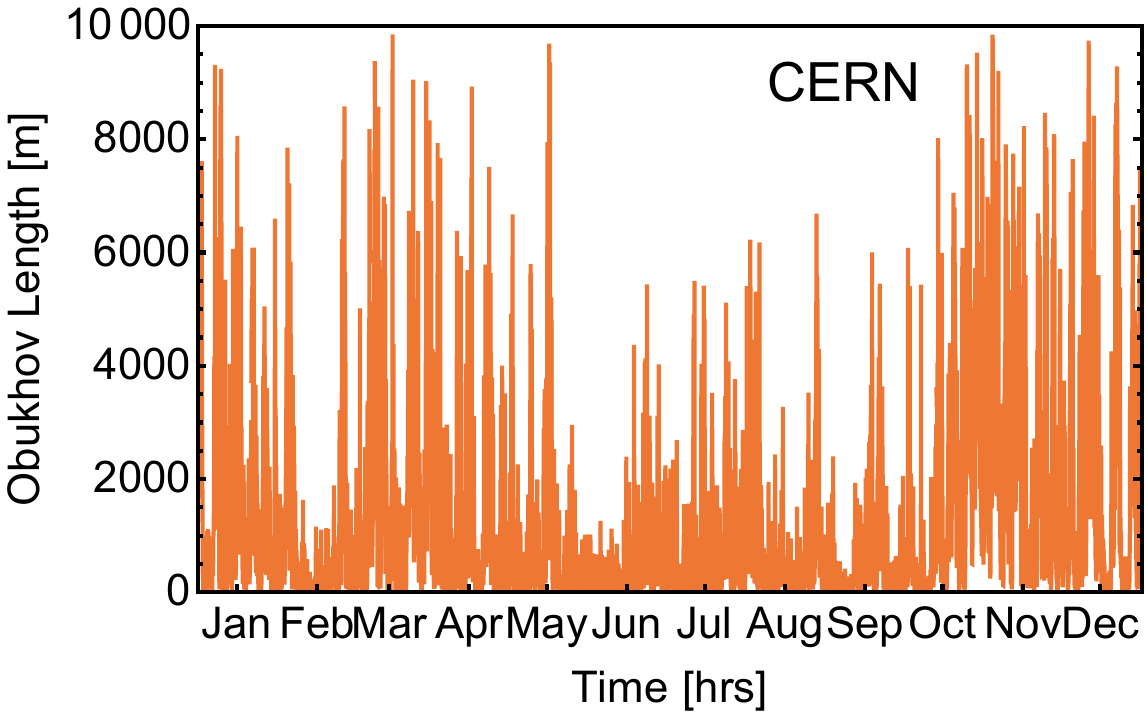}
    \includegraphics[width=0.32\textwidth]{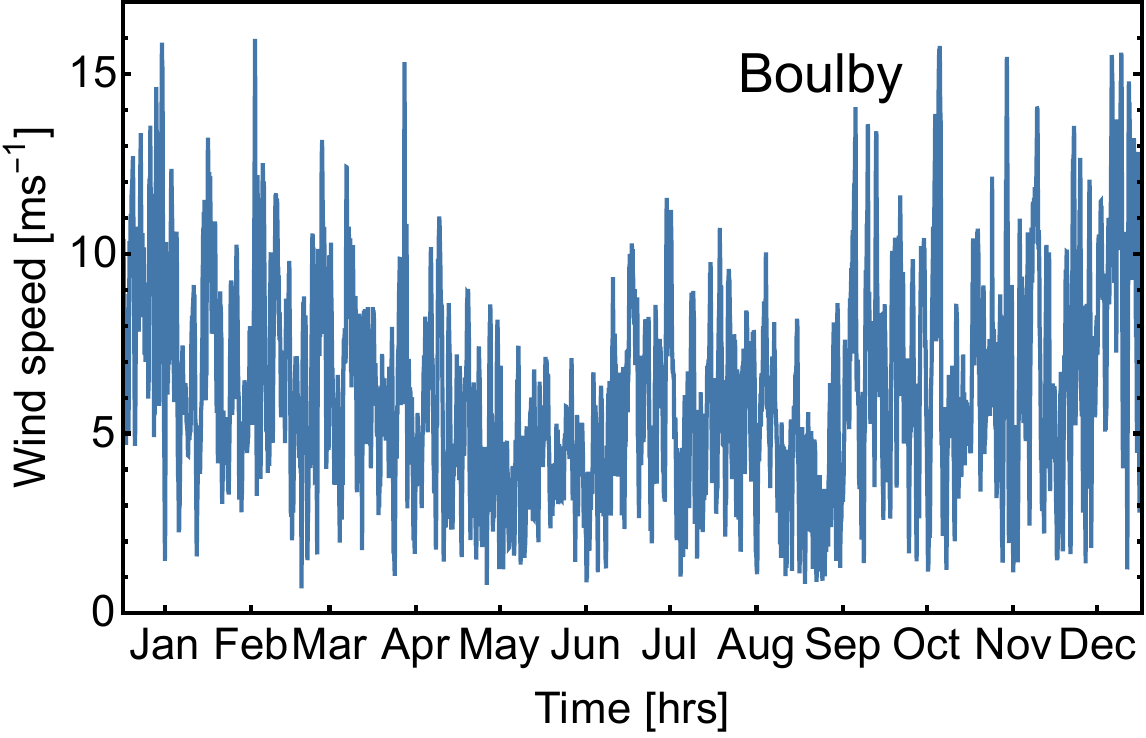}
    \includegraphics[width=0.32\textwidth]{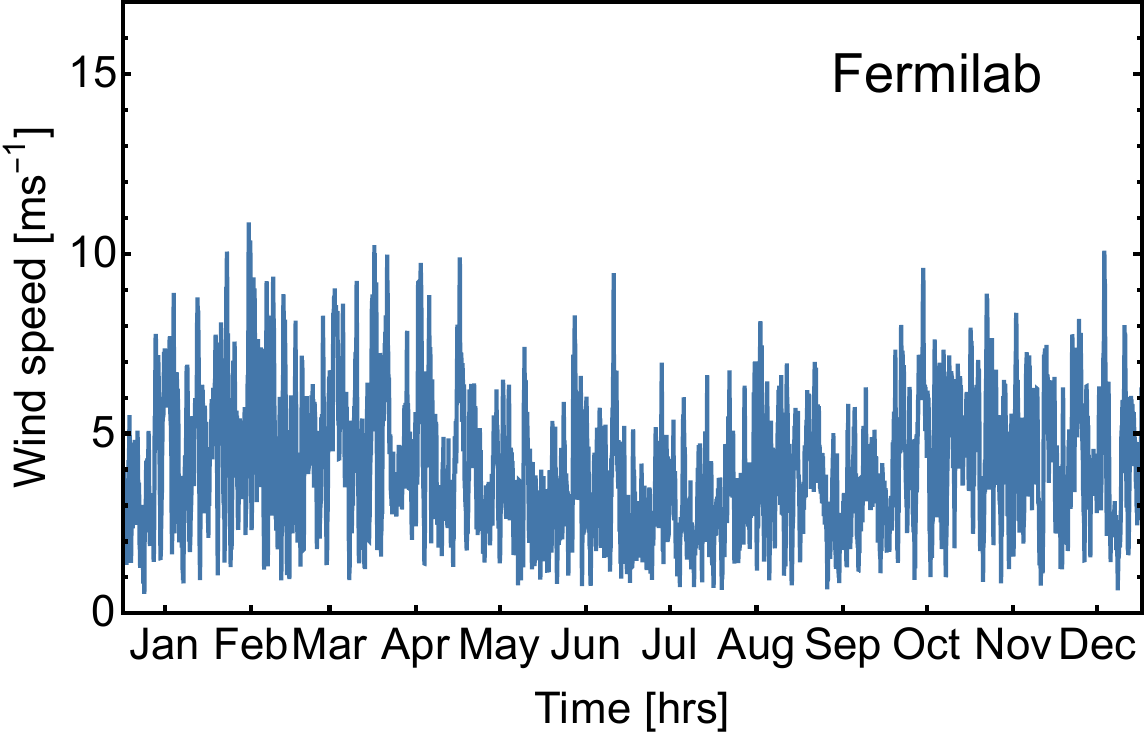}
    \includegraphics[width=0.32\textwidth]{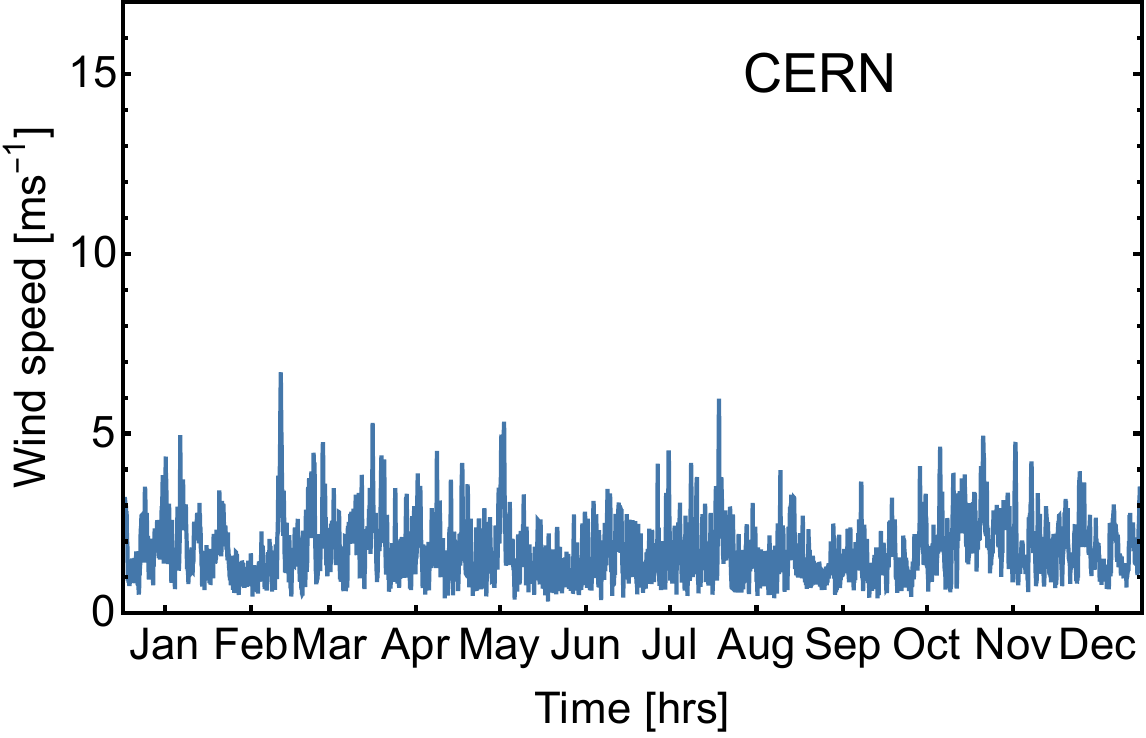}
    \includegraphics[width=0.32\textwidth]{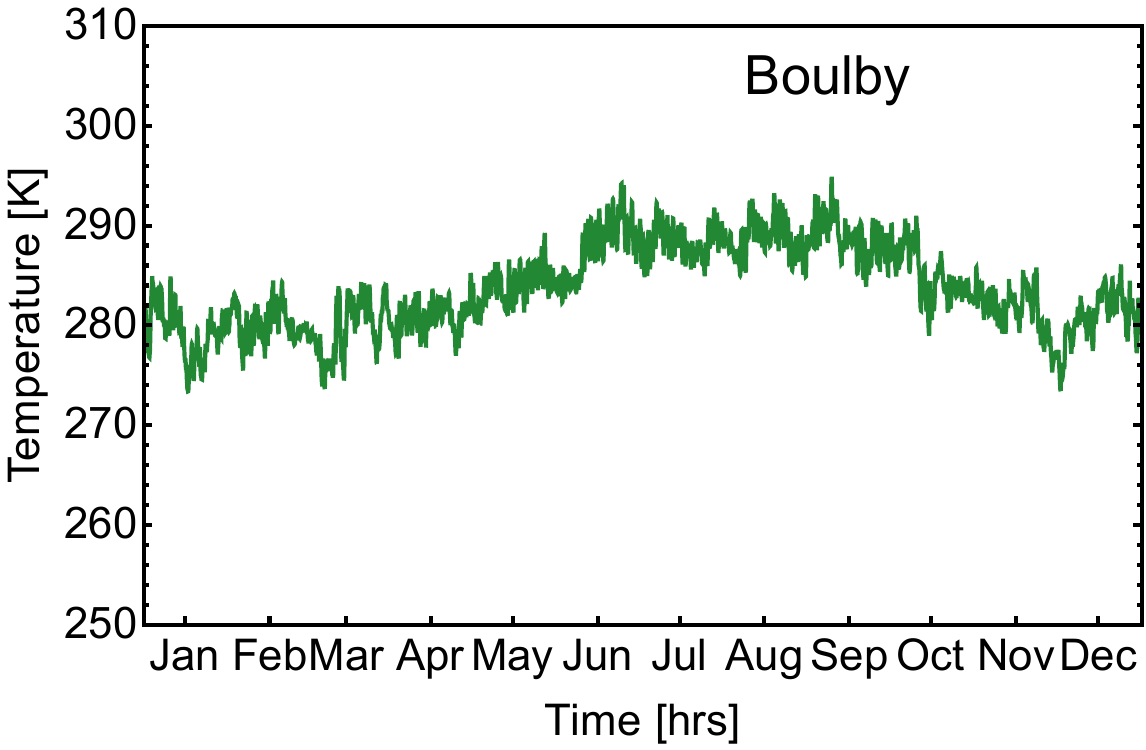}
    \includegraphics[width=0.32\textwidth]{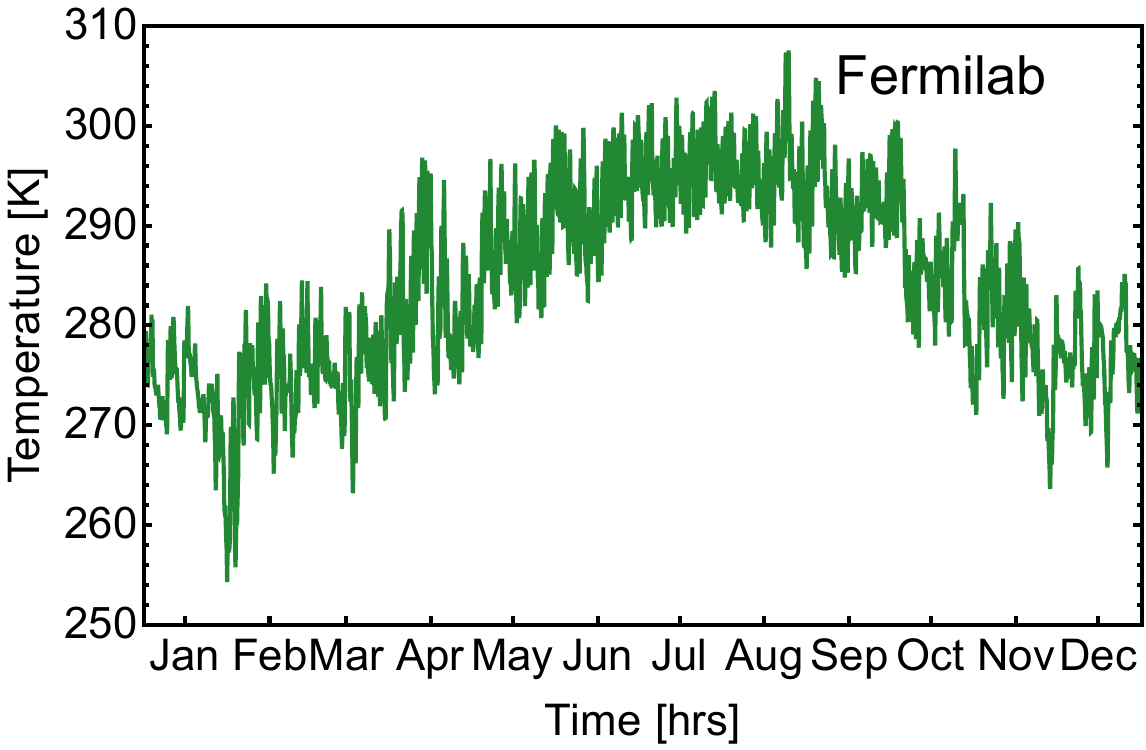}
    \includegraphics[width=0.32\textwidth]{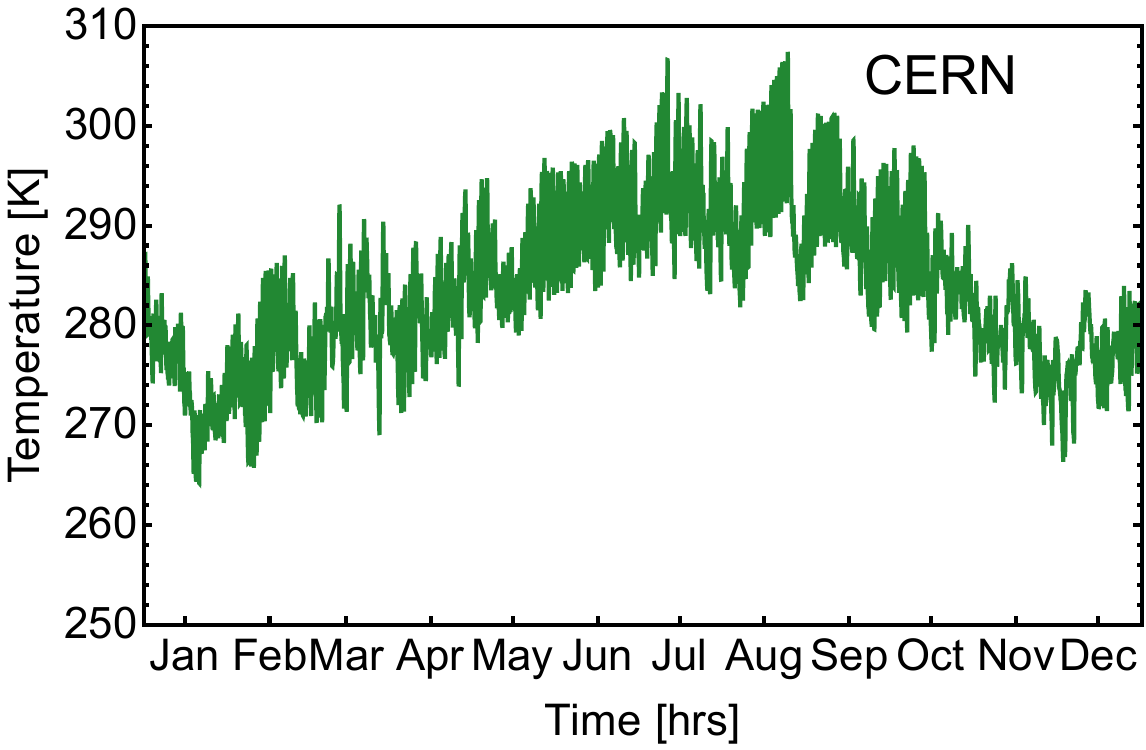}
    \includegraphics[width=0.32\textwidth]{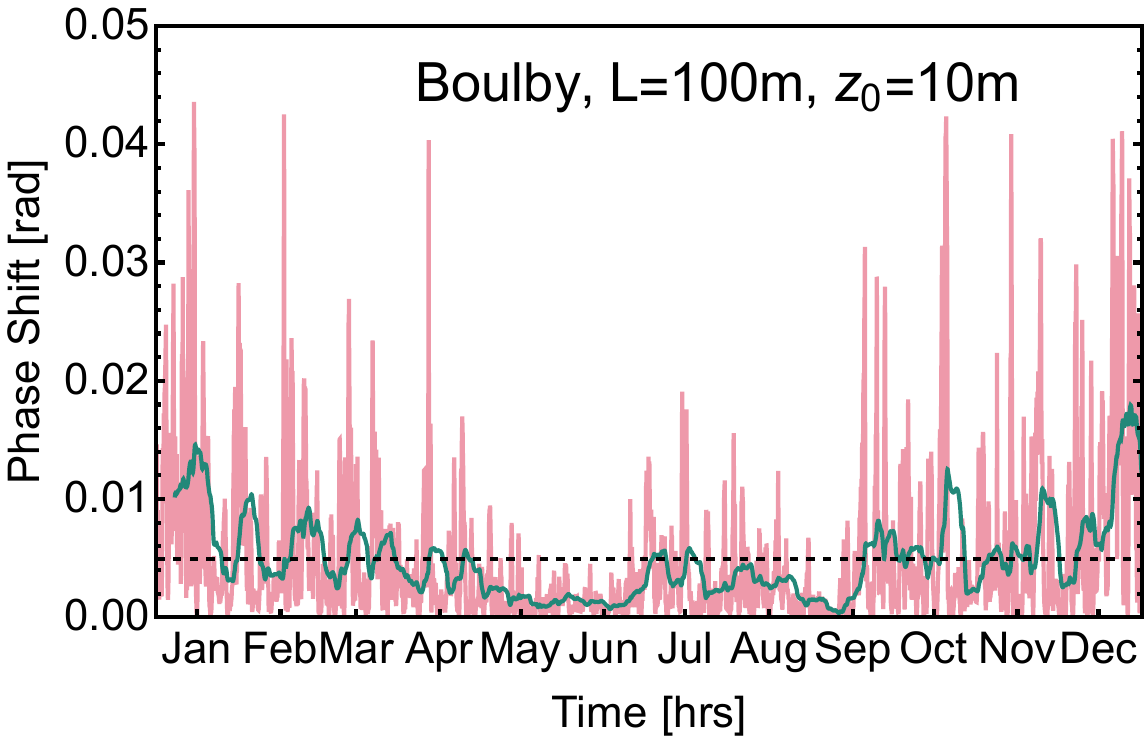}
    \includegraphics[width=0.32\textwidth]{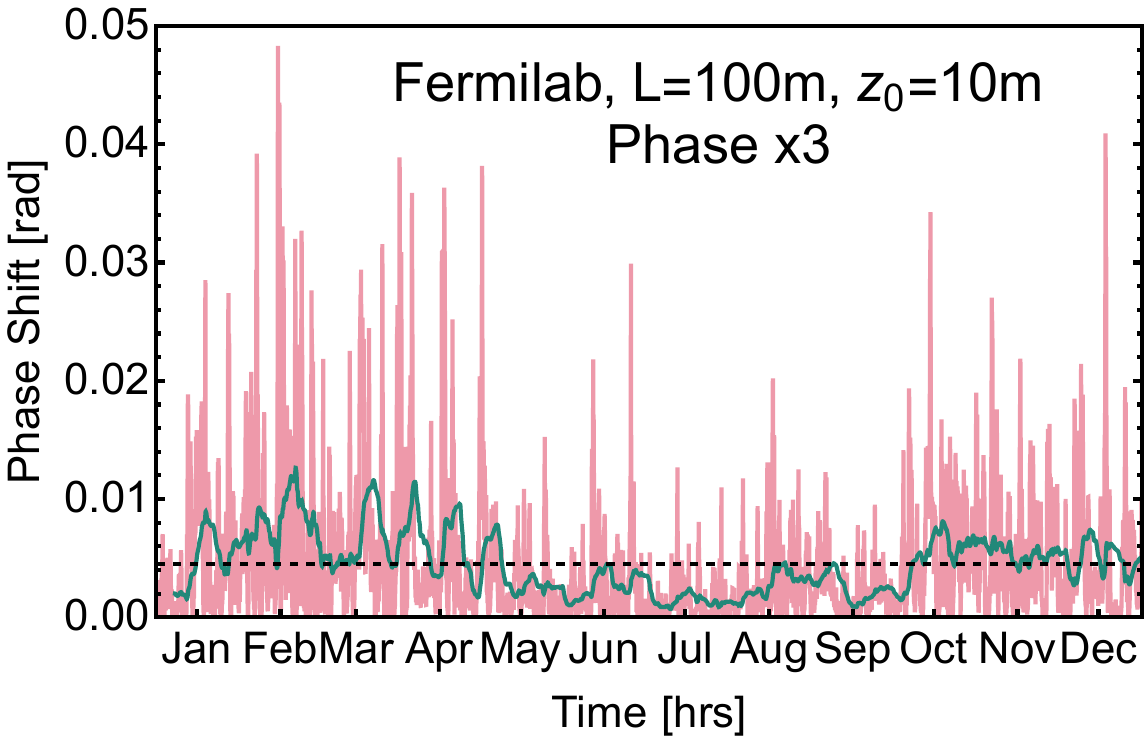}
    \includegraphics[width=0.32\textwidth]{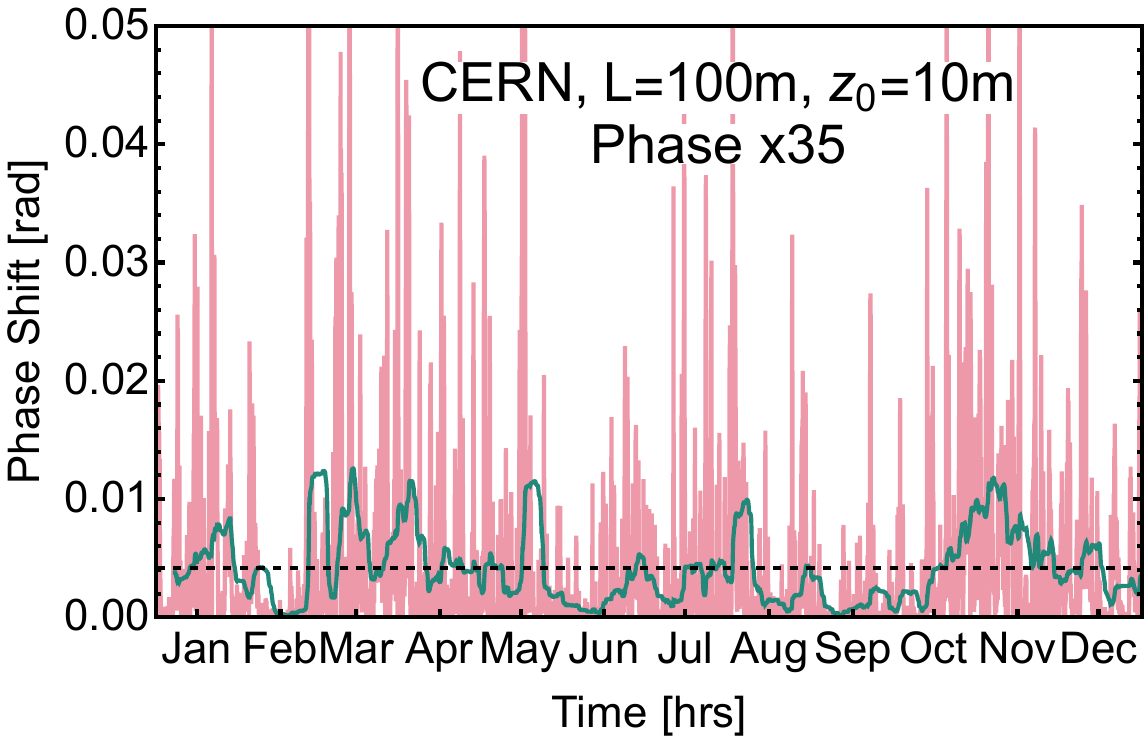}
    \caption{Comparison of atmospheric parameters and temperature phase noise modeled over a year (2023) for Boulby mine, Fermilab and CERN. The first row of plots show a comparison of Obukhov length, the second show wind speed, the third ambient temperature and finally phase noise. The black dashed lines in the phase plots show the mean of phase across the whole year while the green lines show a weekly moving average. The Fermilab and CERN phase noise has been multiplied by a factor of 3 and 35 respectively to better compare the average of level of noise between sites. Data from the ERA5 database~\cite{ERA5}.\\}
    \label{fig:site-compare}
\end{figure*}

To analyze how atmospheric temperature noise affects the detector response,
we use hourly data over a year of measurements (2023) from the ERA5 database~\cite{ERA5} to model key parameters entering Eq.~\eqref{eq:sg2}.
The data was sampled from the average of a  $1^\circ$ latitude by $1^\circ$ longitude area on the surface of the Earth ($\sim\SI{100}{km}\times\SI{100}{km}$) surrounding Boulby mine in north-east England, Fermilab in the midwest of the USA, and CERN on the France-Switzerland border - three proposed locations for $\SI{100}{\metre}+$ scale AI experiments. Our model depends on three key parameters: the outer scale of turbulence which we equate to the Obukhov length $L_O$, the mean wind speed $U$, and ambient temperature $T_0$. Fig.~\ref{fig:site-compare} shows each of these parameters over a year of data taking with a resulting phase time series calculated from our model of atmospheric temperature noise. Each plot is shown side by side on the same axis to compare the locations. 

Each parameter impacts the noise spectra in a different way: the Obukhov length determines the physical extent of turbulent structures that can form, the wind speed determines the rate that the turbulent eddies are advected past the detector, and the ambient temperature determines the deviation of temperature fluctuations from the mean. From Eq.~\eqref{eq:obukhov}, we see the Obukhov length also linearly depends on the ambient temperature, complicating the relationship between these parameters. Regardless, the ambient temperature appears to have the least significant effect on the noise compared to the other parameters. The Obukhov length has a minimal impact on the noise above a certain height -- i.e., when $L_O = \Lambda \gg k$ in Eq.~\eqref{eq:GT}. For shorter lengths, the outer scale matters much more, effectively restricting the production of turbulent eddies. We can thus interpret the results in Fig.~\ref{fig:site-compare} as the noise level being generally set by the average wind speed but restricted at times by the Obukhov length. The spikes in $L_O$ correspond with the spikes seen in the noise, but the height of the noise spike is determined by the average wind speed.

The wind speed is the overall dominant factor in determining noise level, and may be used as a simple predictor of the suitability of different sites to house AI experiments. This is unsurprising, as a higher wind speed is indicative of more temperature perturbations being advected past the detector over time, and thus subjecting it to more noise. In~\cite{Brundu:2022mac}, analytic approximations of the temperature noise spectrum show that under Taylor's frozen turbulence hypothesis that the amplitude of the temperature noise spectrum scales with average wind speed as a power law. 

Overall, Boulby appears to be a significantly noisier location for an AI experiment. The phase noise for Fermilab and CERN has been multiplied by a factor to compare the overall average noise (black dashed lines) and weekly moving average (green lines). The noise level at Boulby is approximately three times greater than Fermilab and 35 times greater than CERN.  

While Fermilab and CERN appear to have a larger variation in temperature, the average wind speeds in general are greater for Boulby, which is the dominant factor in determining the noise level. Despite recording the longest Obukhov lengths for the CERN site, the low average wind speed ultimately renders the atmospheric temperature GGN negligible for this site compared to the others. The Boulby site borders the North Sea where we would generally expect more turbulent conditions and greater variability in wind speeds than over land. 

However, throughout the year the noise level changes dramatically, particularly around May and June. The mean noise (denoted by the black dashed line), and moving average (in green), show that while there are large spikes in noise, there are plenty of times throughout the year where the noise level is an order of magnitude lower than the maximum. It is also clear to see the highly variable and time dependent nature of atmospheric temperature noise in this experiment, which may require continuous active monitoring to fully mitigate. In addition to being a useful analysis for comparing potential experiment sites, these calculations also show how running the interferometers may be more viable at certain times of year. 

The definition of the outer scale of turbulence $\Lambda$ is debated and would most accurately be found by measuring it at any particular location and time. Roughly, we expect it correspond with height $z$ above the ground, and thus the expected size of turbulent structures that can form in the atmosphere~\cite{Brundu:2022mac}. As we consider only effects in the surface layer up to a height $L_O$, this seems like a natural approximation for the outer scale. Other studies have shown the outer scale explicitly depends on both $z$ and $L_O$ with additionally numerical factors~\cite{Tofsted_2000}. With this and the minimal impact the outer scale of turbulence has on the noise above a few tens of meters, we justify using the Obukhov length as an approximation of the outer scale.


\begin{figure*}[p]
    \centering
    \includegraphics[width=0.8\textwidth]{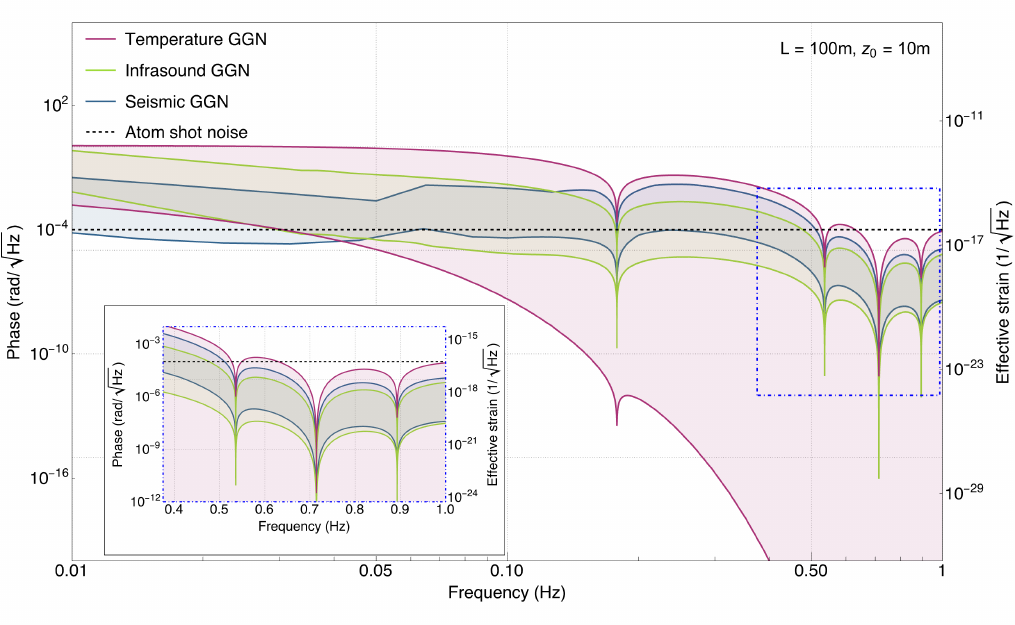}
    
    \includegraphics[width=0.8\textwidth]{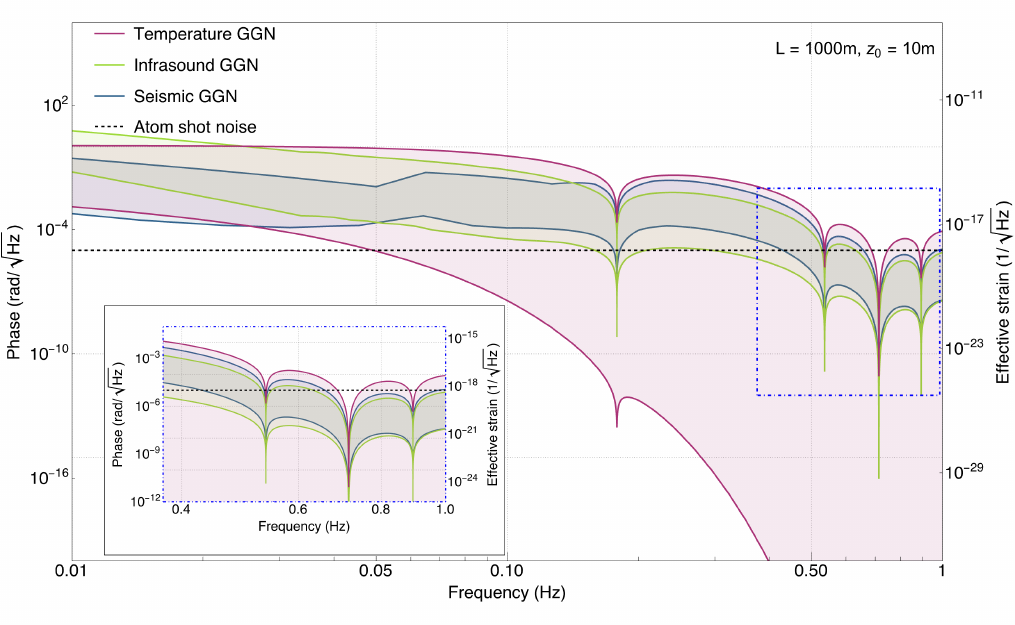}
    
    \caption{Estimates of global upper and lower bounds for seismic, atmospheric pressure and atmospheric temperature GGN in vertical atom interferometers. A $\SI{100}{\metre}$  and $\SI{1}{\kilo\metre}$ baseline experiment are shown on the top and bottom respectively. The black dashed line shows the expected level of atom shot noise in the experiment as stated in Tab.~\ref{tab:params}. The blue dot-dashed box shows a reduced region plotted as an inset in each plot.}
    \label{fig:global}
\end{figure*}

\begin{figure*}[p]
    \centering
    \includegraphics[width=0.8\textwidth]{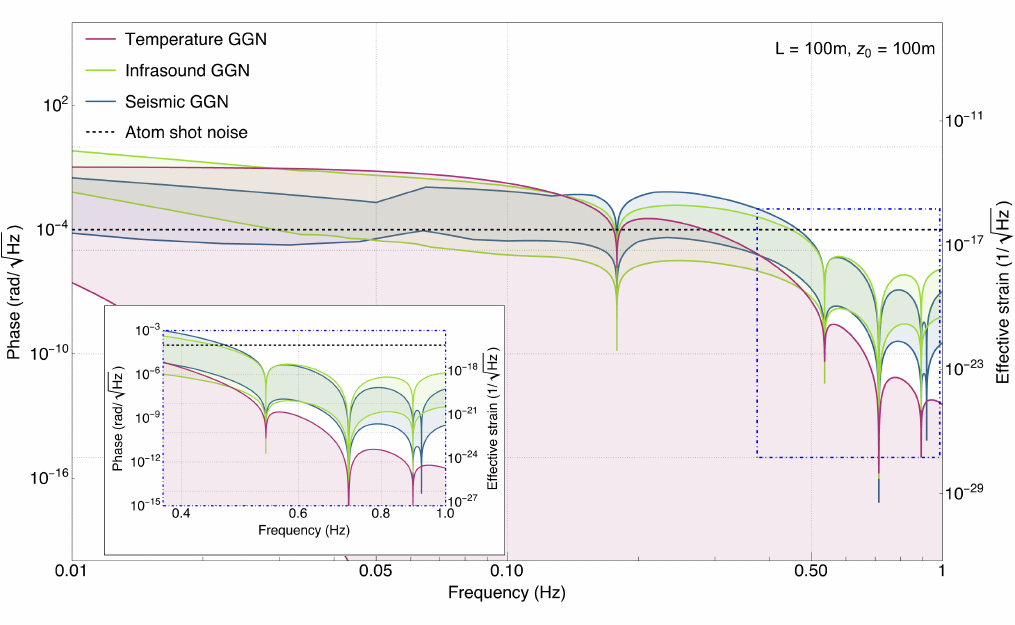}
    
    \includegraphics[width=0.8\textwidth]{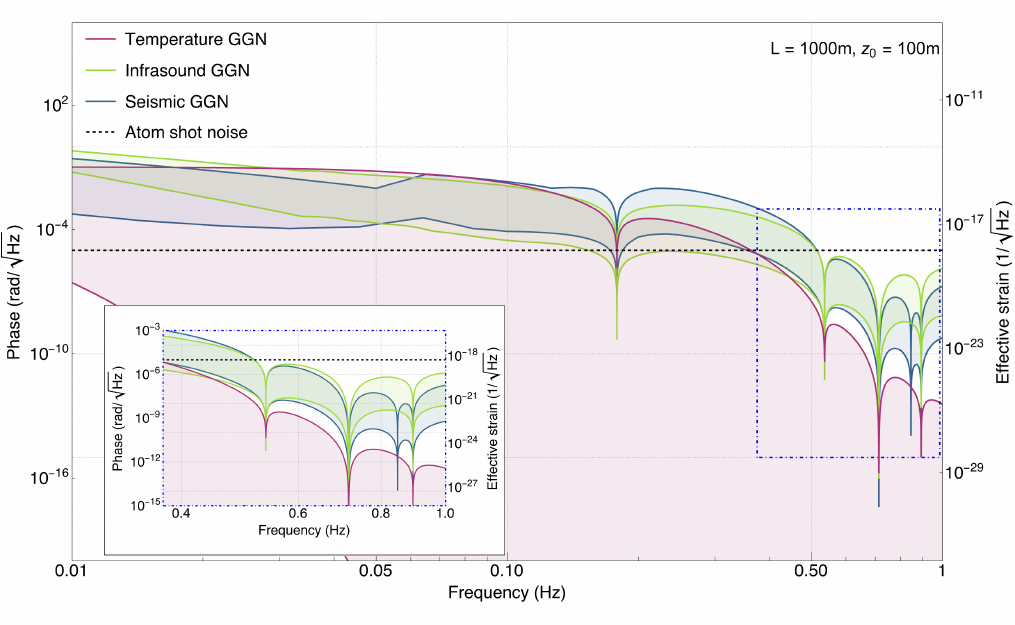}
    
    \caption{Estimates of global upper and lower bounds for seismic, atmospheric pressure and atmospheric temperature GGN in vertical atom interferometers. Identical to Fig.~\ref{fig:global} except each experiment is located $\SI{100}{\metre}$ underground.}
    \label{fig:globalr100}
\end{figure*}


\subsection{Comparison with seismic noise}
\label{subsec:compare-seismic}

\begin{table}[t]
    \centering
    \begin{tabular}{c c c c} 
    \toprule
     & $\Lambda$ [m] & $U$ [ms$^{-1}$] & $T_0$ [K] \\ 
    \midrule
    Maximum & 6600 & 30 & 300 \\ 
    \hline
    Minimum & 0.07 & 0.7 & 250 \\ 
    \bottomrule
    \end{tabular}
    \caption{Parameters used in our calculation of atmospheric temperature noise spectra. Listed are the maximum and minimum case of outer scale $\Lambda$, mean wind speed $U$, and ambient temperature $T$. Data from the ERA5 database~\cite{ERA5}.}
    \label{tab:noise_params}
\end{table}

Deriving global models for atmospheric temperature noise is less concrete than the ambient infrasound models from a network of sensors. Instead we employ our derived model of temperature noise assuming Taylor's hypothesis and the Greenwood-Tarazano spectrum. Outside of the atom interferometer parameters we assume, our model of atmospheric temperature noise relies on three parameters: the outer scale of turbulence $\Lambda$, mean wind speed $U$, and ambient temperature $T_0$. We can then use approximate global modeling of these parameters to examine the highest and lowest typical values for noise in an atom interferometer. Each of these parameters may vary across the Earth's surface on short timescales, making establishing robust global models a challenge. However, our model at least confines the possible range of noises. The ERA5 database~\cite{ERA5} provides the data we have sampled to determine the upper and lower values for noise, with their values listed in Table~\ref{tab:noise_params}. 

We now compare the impact of noise from the derived models of atmospheric pressure and temperature noise in this work with the established Peterson seismic noise models~\cite{Peterson:1993}. The leading seismic source of density fluctuations is the fundamental Rayleigh wave mode. Rayleigh waves are surface waves traveling between media of different densities. They are characterized by their surface displacement spectra. Using the Peterson high- and low-noise models, which are global composite fits to an international seismic network, as the input displacement spectra and the previously derived semiclassical seismic phase response of a vertical gradiometer AI we compute the estimated phase noise spectrum assuming typical parameter values~\cite{Badurina:2023}. The seismic phase noise depends on three parameters: the Rayleigh wave velocity which we take to be \SI{205}{\meter\per\second}, the density of the ground $\rho = \SI{1380}{kg\per\meter\cubed}$, and the material properties of the local surface characterized by the Poisson ratio $\nu = 0.27$. Seismic data recorded on seismometers can include signatures of infrasound and temperature fluctuations in the lower frequencies. We leave disentangling these effects for future work.

Figures~\ref{fig:global} and~\ref{fig:globalr100} show a comparison between temperature, infrasound, seismic GGN and atom shot noise. The figures shows the full range of noise spectra, from \SIrange{0.01}{1}{Hz}, while the inset plots show a finer range of frequencies, as highlighted by the blue dot-dashed box. The upper plot of Fig.~\ref{fig:global} shows the noise spectra assuming the parameters in Table~\ref{tab:params} for a baseline of $\SI{100}{\metre}$, while the lower plot assumes a baseline of $\SI{1}{\kilo\metre}$. The same format is repeated in Fig.~\ref{fig:globalr100}, but $z_0 = \SI{100}{\metre}$ instead of $z_0 = \SI{10}{\metre}$. The noise in both figures is compared to the expected level of atom shot noise. 

As Fig.~\ref{fig:global} demonstrates, for both a $\SI{100}{\metre}$ and $\SI{1}{\kilo\metre}$ experiment, the high noise GGN models dominate at almost all frequencies up to \SI{1}{Hz}. It is interesting to note how closely the noise models are to each other. With the exception of the low noise temperature model, each source of GGN results in approximately the same impact in the detector. There is much greater variability in the temperature GGN models due to the highly time-dependent nature of the noise. Pressure and seismic GGN are characterized by a fixed spectrum at each location on Earth, limited by the high- and low-noise models. While a specific location on Earth may have a typical temperature-noise profile, the ever turbulent nature of the atmosphere ensures the noise varies in this broad range. Under ideal conditions, i.e., low wind and turbulent eddy production, this noise may only be an issue at frequencies below $\sim \SI{0.05}{\hertz}$.

The left axis of each plot in Fig.~\ref{fig:global} shows the phase noise in $\mathrm{rad}/\sqrt{\mathrm{Hz}}$ while the right axis shows the effective strain to connect to noise in laser interferometer experiments. The effective strain is defined using the atom interferometer transfer function, which introduces the sudden dips in noise sensitivity seen in the plots. Taking just the magnitude of the conversion between phase and strain we find the effective strain spectrum
\begin{equation}
\begin{split}
    S_h^\mathrm{eff} &= \frac{S_\phi}{2n\kappa L},\\
    &\approx 5\times 10^{-13} S_\phi \left(\frac{n}{1000}\right)^{-1}\\
    &~~~~~~\times\left(\frac{\kappa}{9.002\times 10^{6}\;\mathrm{m}^{-1}}\right)^{-1}\left(\frac{L}{\SI{100}{\metre}}\right)^{-1}.\\
\end{split}
\end{equation}

As we have shown in Figs.~\ref{fig:depth_gradphaseshift} and~\ref{fig:infra-gradstrain}, the impact of pressure and temperature GGN may be partially mitigated by placing a detector further underground. Therefore, in Fig.~\ref{fig:globalr100}, we show the gradiometer phase for seismic, pressure and temperature GGN for a gradiometer depth of $z_0 = \SI{100}{\metre}$.
At higher frequencies, the noise level for each noise source drops dramatically, with atom shot noise dominating for $\sim$\SIrange{0.5}{1}{\hertz}. Depth appears particularly effective for mitigating atmospheric temperature noise as seen from comparing the upper and lower plots of Fig.~\ref{fig:depth_gradphaseshift}. However, at low frequencies, the gradiometer depth appears to do little to mitigate any source of GGN. An interesting feature appears in the seismic GGN spectra at this increased depth. In addition to the dips in sensitivity from the AI transfer function, the seismic GGN curve has an additional depth-dependent feature about $\SI{0.9}{\hertz}$, resulting from the model's dependence on the difference of two exponential functions with different exponents depending on depth. This results in new zeroes of the function at different depths~\cite{Mitchell:2022,Badurina:2023}.

\section{Discussion and implications} \label{sec:discussion}

We now discuss the results from the previous section, how the noise characterization may be improved in future work, and potential mitigation strategies to combat atmospheric GGN. We emphasize that our key results are found in Sec.~\ref{sec:atmo-ggn} where we extend previous work to model the accelerations induced by pressure and temperature noise. These noise models are experiment independent, directly translating to a strain measure in laser interferometer experiments and can be simply converted to an atom interferometer phase shift with a sequence specific transfer function. 

\subsection{Atmospheric modeling challenges}

Working in the spectral domain allowed us to characterize atmospheric GGN through well-established analytical methods. This approach implicitly assumes that the atmospheric sources are stationary and Gaussian. The atmosphere is a complex dynamical system that contains domains of intermittency, specifically on the scales that cause gravitational noise in our detection band~\cite{Schertzer:1989}. This leads to divergence from stationarity on short- and mid-term scales, but tends to average out over longer duration measurements on the order of a few years.

Our assumption of Gaussianity means the building blocks of the amplitude spectral densities are simply mapped between the random atmospheric perturbation's amplitudes and their Fourier modes. This makes the underlying statistics analytically tractable. In scenarios where the underlying process of fluctuations is non-linear, the statistics become more complicated and thus are not exactly captured by Gaussian statistics, and when transforming to the Fourier domain, the statistics do not remain invariant. This leads to polyspectra in the Fourier domain and requires deeper statistical analysis to extract and map the resulting Fourier amplitudes of the random atmospheric sourcing field~\cite{Brillinger:2012}. For active search campaigns, continuous monitoring of the key parameters that affect the modeling in the time domain would provide proper reconstruction and noise filtering.

These considerations extend to the difficulties in modeling atmospheric temperature perturbations from the spectral domain. The models we use to determine turbulent production and advection of temperature eddies primarily come from empirical measurements. Consequently, determining the exact response of AIs to temperature noise sources is challenging, given that they are expected to be more sensitive instruments than those used to propose atmospheric models initially. However, this suggests AIs could serve as useful tools for improving atmospheric modeling, in turn enhancing noise mitigation strategies.

In Sec.~\ref{subsec:site-compare} we explored the variability of temperature noise throughout a year of measurements. A limiting factor in using ERA5 data is that it gives hourly measurements of atmospheric parameters and estimates them over large areas. We take a relatively large area of data ($1^\circ$ latitude by $1^\circ$ longitude) across a year to calculate an approximate level of noise. However, to discern the spectral features of the noise in the frequency range we are interested in would require atmospheric measurements sampled on the order of $T\sim \SI{1.4}{\second}$, and site-specific measurements to improve their accuracy and account for geographical variations.

The noise level depends somewhat on our assumptions about Taylor's frozen turbulence hypothesis and the Greenwood-Tarazano spectrum of turbulence. While these assumptions were sufficient for this work, site-specific atmospheric measurements would enable the development of a more refined model in future studies.

\subsection{Mitigation strategies}

We now consider potential mitigation schemes for atmospheric GGN. The primary goal of any GGN mitigation is to keep the test mass (the atoms) far from the source of gravitational perturbations and to actively monitor the gravitational environment around the AI. Our frequency range of interest is below \SI{3}{Hz} for tests of fundamental physics with AIs. At these frequencies, infrasound perturbations transported at the speed of sound have wavelengths on the order of 100s of meters to 10s of kilometers. For these wavelengths, obstacles such as buildings provide little to no attenuation. The pressure wave is attenuated by a factor $\exp(-\gamma(\omega) d)$ where $d$ is the wall thickness and $\gamma(\omega)\propto \omega^2$ is the attenuation coefficient. Passive structures would need to be carefully designed to suppress the low-frequency infrasound considered here. As shown in Sec.~\ref{sec:atmoggn-results}, moving the AI deep underground may help but is still limited in effectiveness at the lowest frequencies. A possible recourse is to implement high-resolution active infrasound monitoring for post-correction or real-time filtering.

While infrastructure may have limited impact on the propagation of infrasound, the advection of temperature perturbations relies on the unimpeded path of the wind. Thus, the deliberate placement of obstacles above and around the location of an AI experiment may prove to be a powerful tool for passive noise mitigation. Depth also appears to be an effective passive mitigation strategy, suppressing the noise level for temperature GGN more effectively than infrasound or seismic noise at higher frequencies. This is because temperature perturbations are restricted to be entirely above the ground. However, with these atmospheric effects suppressed, other sources of temperature noise may become more dominant with depth, such as turbulence in the air surrounding the experimental setup. This warrants further investigation to assess the viability of placing vertical long-baseline AI experiments underground.

Multi-gradiometry techniques may also be a powerful tool for mitigating atmospheric pressure and temperature GGN~\cite{Mitchell:2022, Badurina:2023}. Studies optimizing the configuration of several atom sources along a vertical baseline show that it is possible to characterize seismic GGN across the detector. This recovers several orders of magnitude in sensitivity, particularly from $\SI{1}{\hertz}$ down to $\SI{0.1}{\hertz}$, reducing the noise below atom shot noise level. The pressure wave nature of infrasound and similar noise level found from the global models suggest this would also be an effective strategy for atmospheric pressure noise mitigation. Temperature noise would likely also benefit from this strategy. The vertical fountain configuration we explore in this work ensures that the noise is entirely coherent along the baseline, greatly aiding the characterization of these noise sources which primarily travel horizontally.

While these passive strategies cannot entirely mitigate GGN and have limited effect at frequencies below $\SI{0.1}{\hertz}$, combining them with continuous active monitoring of the environment surrounding an experiment offers a path forward. This can be achieved by deploying very broadband seismometers, tilt-sensors, low-frequency infrasound sensors, and temperature sensors that each have ultra-low self-noise in the frequency bands of interest. These instruments also need high sampling rates and bandwidth with minimal digital noise to achieve near real-time feedback control. Such instrumentation is available commercially at reasonable cost. A known limiting factor for these instruments, e.g., broadband seismometers, is their self-noise (thermal and electronic) dominating below 10s of mHz as well as their sensitivity to ground tilt in the low-frequency regime~\cite{Ringler:2010,Lin:2022}. This is one area where future technological advancement is expected to benefit AIs directly, but our current resources can already be of great help in mitigating noise.

\subsection{Dependence on AI sequence }

We employed one example of an atom interferometer measurement sequence in our characterization. In general, the phase shift depends on the trajectory geometry and the pulse sequencing. Both factors can change the location of the atoms in the gravitational potential and thus alter the interferometer arm path lengths and phase accumulation. We limited ourselves in this study to a fountain configuration where the atoms follow a parabolic trajectory and took the short light-pulse duration limit. This means we treat each atom-light interaction as instantaneous, easing the description of the atom trajectory while in the gravitational and optical potential. In reality, the light-pulse times can range from 100s of \si{\nano\second} to 100s of \si{\micro\second}, and for more robust atom-optics sequences in the range of a few \si{\milli\second}, altering the derived phase shifts.

The large momentum transfer was treated in a symmetric way, with light pulses used to increase the space-time area of the interferometer via counter-propagating directions acting symmetrically on both arms, which is captured by the linear scaling of LMT order. The overall time-dependent phase noise caused by atmospheric effects was considered in the absence of other DC gravity gradient effects that would lead to a bias of the overall phase shift. These static effects have been investigated previously~\cite{Schilling:2020}.

Other atom interferometer geometries different from the Mach-Zehnder sequence would require recalculation, as the response functions are unique to each configuration, but could still make use of the fluctuating potentials discussed here. Other typical atom interferometer sequences include Ramsey-Bord\'e sequences\footnote{These sequences are the geometry required for precise atom interferometric measurement of the fine structure constant.}~\cite{Muller:2009} and multi-loop configurations\footnote{Multi-loop configurations are repeated exchanges of the interferometer arms forming multiple diamonds leading to highly sensitive resonant measurements proposed for narrowband searches of GW and cancellation of systematic effects such as Coriolis acceleration phase shifts.}~\cite{Graham:2016}. 
Correlating data from multiple AI along the baseline has also been proposed to minimize GGN~\cite{Mitchell:2022, Badurina:2023}. Different vertical AI sequences are expected to have broadly the same response to atmospheric sources of GGN, with the main difference entering through frequency-dependent features of the transfer function $T_\phi$.

\section{Conclusion and Outlook}\label{sec:conclusion}

Atom interferometers promise to be powerful tools for making measurements in many areas of fundamental physics.
This work presents the first comprehensive characterization of atmospheric gravity gradient noise (GGN) for vertical atom interferometers, 
an important step forward in our understanding of noise sources for future detectors.

While previous studies examined atmospheric pressure GGN for horizontal configurations, a characterization for vertical interferometers necessitated a novel analysis.
These setups experience correlated noise along their baselines rather than uncorrelated noise between test masses. 
Additionally, atmospheric temperature noise has previously not been studied for AI experiments in either horizontal or vertical configurations.

We have developed robust mathematical frameworks for modeling both pressure and temperature-induced GGN. 
For pressure noise, we extended existing infrasound wave models to characterize acceleration fields surrounding vertical detectors. 
For temperature noise, we worked in the spectral domain, implementing an empirically-derived wavenumber spectrum optimized for the low-frequency regime relevant to atom interferometry. 
This analysis demonstrated that vertical configurations offer inherent advantages over horizontal setups, as the coherent subtraction of noise along the baseline provides better noise rejection compared to the incoherent addition typical in horizontal experiments. 
For both noise sources, the AI closer to the surface will dominate the noise profile.

Our investigation of depth dependence revealed distinct behaviors for different noise sources. 
Temperature-induced GGN showed strong suppression with depth, as thermal eddies are exclusively advected above ground. 
In contrast, pressure-induced GGN exhibited only exponential decay below the surface, making it more challenging to passively mitigate through underground placement. 
Both sources proved particularly difficult to suppress at frequencies below \SI{0.1}{Hz}, highlighting a key challenge for future experiments.

By combining our theoretical framework with global infrasound monitoring data, we established the first comprehensive bounds on expected atmospheric GGN levels for vertical atom interferometers. We utilized measurements from the International Monitoring System to define pressure noise limits and developed novel approaches for bounding temperature noise using ERA5 database parameters. 
We examined a case study comparing potential experimental sites at Boulby mine, Fermilab, and CERN, which demonstrated significant temporal and geographic variability in noise levels. 
The coastal proximity of Boulby is likely to be a key factor in this site, exhibiting a higher overall noise level with a higher average wind speed.

This work has several important implications for future experiments.
 First, our analysis suggests that atmospheric GGN poses challenges comparable to seismic GGN, particularly below \SI{0.5}{Hz} where both noise sources exceed typical atom shot noise levels. 
 Second, we found that placement at depths of \SI{100}{m} or greater can effectively mitigate noise above \SI{0.5}{Hz}, though low-frequency noise remains a persistent challenge. 
Finally, our results emphasize the importance of continuous environmental monitoring, as atmospheric noise, particularly temperature GGN, exhibits strong temporal variations that must be actively tracked and compensated.

Looking ahead, several promising directions emerge for future research. 
Development of advanced noise rejection techniques, particularly multi-gradiometry configurations optimized for atmospheric GGN, could provide additional sensitivity improvements. 
Further investigation of the interplay between different noise sources at depth, especially the role of local temperature fluctuations in underground facilities, will be crucial for optimizing detector design and placement. 
Additionally, the high sensitivity of atom interferometers to atmospheric effects suggests their potential utility as tools for improving atmospheric modeling, creating a mutually beneficial cycle where better models lead to better noise mitigation strategies.

The tools and frameworks developed in this work provide a foundation for site selection and noise mitigation in next-generation atom interferometry experiments. As these instruments push toward sensitivities exceeding $\delta \phi\sim \SI{1e-4}{\radian/\sqrt{\hertz}}$, particularly at frequencies below $\SI{0.5}{\hertz}$, careful consideration of atmospheric GGN will be essential for achieving their full potential in searches for new fundamental physics.

\section{Acknowledgments}
We are grateful to members of the AION and MAGIS-100 collaborations for comments and productive discussions on this work. We thank John Ellis for comments on the manuscript. We are grateful to S.K. Kristoffersen for providing their infrasound data for modeling our global noise limits.
J.C.\ acknowledges support from a King's College London NMES Faculty Studentship.
T.K.\ acknowledges support from the Gordon and Betty Moore Foundation Grant No. GBMF7945.
C.M.\ is supported by the Science and Technology Facilities Council (STFC) Grant No.\ ST/T00679X/1. 
J.M.\ acknowledges support from the University of Cambridge Isaac Newton Trust and the Science and Technology Facilities Council (STFC) Grant No.\ ST/Y004477/1.

For the purpose of open access, the authors have applied a Creative Commons Attribution (CC BY) license to any Author Accepted Manuscript version arising from this submission. 

The data supporting the findings of this study are available within the paper. 
No experimental datasets were generated by this research. 

\bibliography{References.bib}
\end{document}